\documentclass[
superscriptaddress,  
reprint,
amsmath,amssymb,
aps, pra,
longbibliography,
]{revtex4-2}

\usepackage{graphicx}
\usepackage{dcolumn}
\usepackage{bm}

\usepackage[utf8]{inputenc}
\usepackage[T1]{fontenc}
\DeclareMathAlphabet{\mathcal}{OMS}{cmsy}{m}{n}

\usepackage{braket}
\usepackage{amsmath}
\usepackage{caption}
\usepackage{subcaption}
\usepackage{gensymb}
\usepackage{xcolor}
\usepackage{graphicx}
\usepackage{mwe}
\usepackage{stackengine}

\begin{document}
	
\preprint{AIP/123-QED}


\title[]{Frequency-robust M\o lmer-S\o rensen gates via balanced contributions of multiple motional modes} 

\author{Brandon ~P. Ruzic}
\email{bruzic@sandia.gov}
\affiliation{Sandia National Laboratories, Albuquerque, New Mexico 87185, USA}
\thanks{}
\author{Matthew ~N. ~H. Chow}
\email{mnchow@sandia.gov}
\affiliation{Sandia National Laboratories, Albuquerque, New Mexico 87185, USA}
\affiliation{Department of Physics and Astronomy, University of New Mexico, Albuquerque, New Mexico 87106, USA}
\affiliation{Center for Quantum Information and Control, University of New Mexico, Albuquerque, New Mexico 87131, USA}
\thanks{}

\author{Ashlyn ~D. Burch}
\author{Daniel Lobser}
\author{Melissa ~C. Revelle}
\author{Joshua ~M. Wilson}
\author{Christopher ~G. Yale}
\author{Susan ~M. Clark}
\affiliation{Sandia National Laboratories, Albuquerque, New Mexico 87185, USA}

\date{\today}

\begin{abstract}
    In this work, we design and implement frequency-robust M\o lmer-S\o rensen gates on a linear chain of trapped ions, using Gaussian amplitude modulation and a constant laser frequency. We select this frequency to balance the entanglement accumulation of all motional modes during the gate to produce a strong robustness to frequency error, even for long ion chains. We demonstrate this technique on a three-ion chain, achieving <\,1\% reduction from peak fidelity over a 20\,kHz range of frequency offset, and we analyze the performance of this gate design through numerical simulations on chains of two to 33 ions.  
\end{abstract}

\maketitle

\section{Introduction}


Linear chains of trapped ions are one of the leading platforms for quantum computation in the near term. The application of M{\o}lmer-S{\o}rensen (MS) gates~\cite{sorensen:2000} on these systems has achieved some of the highest two-qubit entanglement fidelities to date, reaching above 99.9\% while targeting the axial motional modes of a two-ion chain~\cite{ballance:2016, gaebler:2016}. To implement powerful quantum algorithms, like digital quantum simulation~\cite{lanyon:2011} and quantum error correction~\cite{shor:1995, kitaev:2003, aharonov:2008}, one must extend these high-fidelity gates to systems of many physical qubits by, for example, increasing the length of the chain and 
individually addressing 
each ion~\cite{debnath:2016, wright:2019}. In this approach, the MS gates provide all-to-all connectivity between ion pairs, but the gate fidelity can suffer due to the residual spin-motion entanglement after the gate in the increased number of spectator motional modes~\cite{sorensen:2000}. 

There have been many successful demonstrations of high-fidelity MS gates by modulating the amplitude~\cite{debnath:2016, wright:2019, zhu:2006, Roos2008, BenhelmRoos2008, choi:2014, tinkey:2022}, frequency~\cite{Leung5ion, wang:2020}, amplitude and frequency~\cite{leung:2018, landsman:2019}, or phase~\cite{green:2015, lu:2019, milne:2020} of the laser beams. These approaches have achieved 97\% to 99.5\% fidelity when targeting the radial modes of a two-ion chain, for which the tighter confinement than in the axial direction allows better cooling, less heating, and faster gates. The modulation techniques improve gate performance by eliminating the residual spin-motion entanglement for ideal experimental conditions and by adding robustness to this quantity in the presence of motional frequency error. For example, simulations of frequency-modulated gates maintain a 99\% fidelity with a motional frequency error of $\pm$1.5~kHz for a two-ion chain~\cite{Leung5ion}, and optimizing over a distribution of gate parameters improves this level of robustness to at least $\pm5$~kHz~\cite{KangBRobust}. 

Nevertheless, motional frequency error remains an important error source in MS gates and their applications. Modulated MS gates attempt to minimize the sensitivity of the residual spin-motion entanglement to frequency error, and as a result, the amount of spin entanglement accumulated during the gate also gains robustness to this error. However, significant errors in the amount of accumulated spin entanglement can remain and create a purely coherent rotation error in spin space, which is especially damaging to the performance of quantum algorithms that involve many gates~\cite{iverson:2020}. This sensitivity to rotation error was recently demonstrated by the repeated application of MS gates with a frequency offset on two-ion and four-ion chains~\cite{wang:2020}.

For longer chains, the sensitivity to frequency error increases due to the higher density of motional modes. Further, the majority of frequency-robust gate designs become more difficult to implement due to more stringent experimental requirements, including the need to account for all modes by linearly increasing the number of optimized pulse-shape parameters with the number of ions~\cite{Leung5ion}. Robust gate designs exist that
reduce this requirement by only targeting closely spaced ions or a reduced set of motional modes~\cite{landsman:2019}, but the experimental requirements to implement these techniques can still grow with longer chains. Modulated gates on longer ion chains can require larger laser powers~\cite{wang:2020} and generally have a higher sensitivity to drift in the calibrated model parameters (e.g.~motional frequencies, ion separation, laser power, and gate duration) that are used during the optimization of pulse-shape parameters~\cite{KangBRobust}.

In this paper, we develop and implement an MS gate with an analytic pulse shape that does not require optimizing a large set of pulse-shape parameters yet is still broadly robust to motional frequency error, even for long ion chains. We perform amplitude modulation during our gate with a simple, Gaussian time dependence that strongly suppresses residual displacement errors in all modes, as long as the detuning from each mode remains sufficiently large. While many studies have demonstrated error suppression using amplitude modulation, including modulation that resembles a Gaussian~\cite{leung:2018, landsman:2019, tinkey:2022}, we also select a specific, constant detuning that balances the amount of entanglement accumulation during the gate from all motional modes and provides robustness to this source of coherent gate error. With the ability to adjust the detuning without significantly impacting displacement errors, we are free to tune the laser frequency to a point where the derivative in the entanglement accumulation with respect to frequency goes to zero. This produces a gate that is first-order insensitive to frequency error, resulting in regions of broad robustness to this error. Our protocol is simple to realize experimentally, as we can optimize performance by calibrating only two pulse-shape parameters: the constant detuning and the peak Rabi rate. 
As a result, our gate design has a low classical computational overhead, facilitating its adoption on other trapped ion quantum processors and making it suitable for systems suffering from moderate amounts of drift. We demonstrate the frequency robustness of our gate on a three-ion chain and analyze this robustness in numerical simulations for chains of up to 33 ions. 

This work is done on the Quantum Scientific Computing Open User Testbed, QSCOUT. We use qubits encoded in the hyperfine clock states of $^{171}$Yb$^+$ ions trapped in a linear chain on a surface trap. Gates are site-selectively driven with an optical Raman transition. Details of the apparatus are described in previous work~\cite{QSCOUTManual}.

\section{Gate Design}

\subsection{MS Gate Model}
\label{sec:model}

We model the application of an MS gate on two ions that are part of a linear chain of ions in a surface trap using the Hamiltonian,
\begin{equation}
\label{eq:H}
H(t) = -\Omega(t)\sum_k S_{y,k} a_k e^{i\delta_k t} + h.c.,
\end{equation}
which is in a rotating frame with respect to the atomic and trap degrees of freedom. The collective spin operator $S_{y,k}$ has the form: $S_{y,k} = (\eta_{1,k}\sigma_{y,1}+\eta_{2,k}\sigma_{y,2})/2$, where $\sigma_{y,j}$ is the $y$ Pauli spin operator for the $j$-th ion targeted by the gate. The Lamb-Dicke parameter $\eta_{j,k}$ can differ for each ion and each motional mode, and $\Omega(t)$ is the Rabi rate of the carrier transition for both ions. In this work, $\Omega(t)$ is a time dependent parameter of the drive field, while $\delta_k$ is effectively held constant in time for each mode. The operators $a_k^\dagger$ and $a_k$ are the raising and lowering operators, respectively, for a harmonic oscillator that represents the motional mode of the ion chain with angular frequency~$\nu_k$. During the gate, a dual-tone laser 
illuminates the ions with detunings $\pm\delta_k=\pm(\delta_\text{c}-\nu_k)$ from their blue and red motional sideband transitions, respectively, where the parameter $\delta_\text{c}$ is the detuning of the blue-detuned laser tone from the carrier transition. For simplicity, we have made the Lamb-Dicke approximation: $e^{i\eta(a_k+a_k^\dagger)}\approx 1+i\eta(a_k+a_k^\dagger)$. We have also neglected the carrier transition and the far-off-resonant sideband transitions with detunings larger than $|\delta_\text{c}|$.

Since the Hamiltonian $H(t)$ acts on each motional mode independently, we can write the propagator $U(t)$ as a product over motional modes:
\begin{equation}
U(t) = \Pi_k U_k(t),
\end{equation}
and the exact analytic solution for $U_k(t)$ is~\cite{sorensen:2000, ruzic:2022},
\begin{gather}
U_k(t) = e^{-i\mathcal{B}_k(t)S_{y,k}^2}D(S_{y,k}\alpha_k(t)), \nonumber \\
\label{eq:U}
\mathcal{B}_k(t)=\frac{i}{2}\int_0^t \left(\frac{\mathrm{d}\alpha_k(t')}{\mathrm{d}t'}\alpha_k^*(t') - \alpha(t')\frac{\mathrm{d}\alpha_k^*(t')}{\mathrm{d}t'}\right)\mathrm{d}t'.
\end{gather}
The displacement operator $D(S_{y,k}\alpha_k(t))=\exp\left[S_{y,k}(\alpha_k(t) a_k^\dagger - \alpha_k^*(t) a_k)\right]$ is conditioned on the spin state of the targeted ions, and $\alpha_k(t)$ describes the phase-space trajectory of the ion chain. The phase $\eta_{1,k}\eta_{2,k}\mathcal{B}_k(t)$, which governs the amount of spin entanglement accrued during the gate, is real and positive (negative) for clockwise (counter-clockwise) trajectories. 

To gain an intuitive picture of the gate dynamics, we express the phase-space trajectory of each motional mode in terms of the parameters of $H(t)$, 
\begin{gather}
\alpha_k(t) = i\int_0^t \Omega(t')e^{-i\delta_k t'}\mathrm{d}t',
\label{eq:alpha}
\end{gather}
where $t=\tau$ corresponds to the end of the gate. From this equation, we see that $\alpha_k(\tau)$ is proportional to the Fourier transform of $\Omega(t)$ evaluated at $\delta_k$, assuming that $\Omega(t)$ is zero before ($t<0$) and after ($t>\tau$) the gate. This is a key insight that will aid our choice of pulse shape for frequency-robust gates, as discussed in section~\ref{sec:pulse_shape}.

In this study, we focus on the robustness of MS-gate performance to a frequency error~$\delta\omega$ that is applied to both laser tones and moves them symmetrically with respect to the carrier transition, resulting in new carrier detunings: $\pm\delta_\text{c}'=\pm(\delta_\text{c} + \delta\omega)$ and sideband detunings: $\pm\delta_k'=\pm(\delta_k+\delta\omega)$, for the blue-detuned and red-detuned tones, respectively. Equivalently, this frequency error can be interpreted as a common change in the motional frequency of each mode: $\nu_k'=\nu_k - \delta\omega$. Although other error sources can affect gate performance, such as laser power fluctuations and anomalous heating~\cite{bruzewicz:2015, boldin:2018}, we choose to focus on frequency error due to the high sensitivity of gate performance to this error~\cite{ruzic:2022}, especially in the context of long ion chains with many closely spaced motional modes. 

\subsection{Performance Metrics}

We use the state fidelity $\mathcal{F}$ as the figure of merit for gate performance, which can be computed by wavefunction overlap:
\begin{equation}
\mathcal{F} = |\braket{\Phi|\Psi(\tau)}|^2,
\end{equation}
where $\ket{\Psi(\tau)}$ is the wavefunction of the ion chain at the end of the gate and $\ket{\Phi}$ is the target state. We assume that we perfectly initialize the ions in the ground spin state $\ket{00}$ and laser cool them to reach the motional ground state $\ket{0}$. The state of the ion chain after the gate is then $\ket{\Psi(\tau)} = \Pi_k e^{i\mathcal{B}_k(\tau)S_{y,k}^2}D(S_{y,k}\alpha_k(\tau))\ket{00, 0}$, and we choose to target the state $\ket{\phi}=1/\sqrt{2}(\ket{00} + i\ket{11})\ket{0}$, a maximally entangled spin state and the motional ground state, as any residual displacement after the gate leads to spin-motion entanglement.

For ideal gate performance $(\mathcal{F}=1)$, we require the propagator at the end of the gate ($t=\tau)$ to take the following form: $U(\tau)=e^{-i \sigma_{y,1}\sigma_{y,2}\theta/2}$, where $\theta$ is the rotation angle of the gate. For our choice of initial and final states, the ideal gate is accomplished for,
\begin{subequations}
\begin{gather}
\label{eq:err_d}
\alpha_k(\tau) = 0 \quad\text{for each $k$},\\
\label{eq:err_r}
\theta = \sum_k \eta_{1,k}\eta_{2,k}\mathcal{B}_k(\tau) =  \pi/2.
\end{gather}
\end{subequations}
Although $\mathcal{F}$ is a sufficient metric for gate performance, we find that it is illustrative in this work to decompose the state fidelity into two contributing terms: displacement error $\epsilon_d$ and rotation-angle error $\epsilon_r$, which arise from inequalities in equations (\ref{eq:err_d}) and (\ref{eq:err_r}), respectively. In the next two sections, we derive the contribution to the state infidelity $1-\mathcal{F}$ from each error separately, where the sum of these errors,
\begin{equation}
    \epsilon_s = \epsilon_d + \epsilon_r,
\end{equation}
is approximately equal to the state infidelity: $\epsilon_s\approx 1 - \mathcal{F}$, for small errors ($\epsilon_d\ll1$ and $\epsilon_r\ll1$). 

\subsubsection{Displacement Error}
\label{sec:err_d}
Displacement error occurs when the coherent displacement at the end of the gate is non-zero, $\alpha_k(\tau) \neq \alpha(0)=0$, and leads to residual spin-motion entanglement. In the phase-space trajectory picture, a gate with no displacement error for a particular mode will produce a closed curve, and any residual displacement (or open curve) contributes to the gate error. We call this contribution the displacement error $\epsilon_{d,k}$ for mode $k$. As accomplished in several previous works, including~\cite{choi:2014, Leung5ion, KangBRobust, milne:2020}, we design gates that are robust to frequency errors for this error mechanism by shaping the laser pulse. 



We derive an expression for the displacement error from each mode by acting the displacement operator for the gate on the initial state and computing the wavefunction overlap with the target state: $\mathcal{O}_{d,k}=\bra{\Phi}D(S_{y,k}\alpha_k(\tau))\ket{00, 0}$. In the spin basis \{$\ket{+}=(\ket{0}+\ket{1})/\sqrt{2}, \ket{-}=(\ket{1}-\ket{0})/\sqrt{2}$\}, the operator $S_{y,k}$ has four eigenvalues: $\lambda_{++}=(\eta_{1,k} + \eta_{2,k})/2$, $\lambda_{+-}=(\eta_{1,k} - \eta_{2,k})/2$, $\lambda_{-+}=(\eta_{2,k} - \eta_{1,k})/2$, and $\lambda_{--}=-(\eta_{1,k} + \eta_{2,k})/2$, corresponding to the spin states $\ket{++}$, $\ket{+-}$, $\ket{-+}$, and $\ket{--}$, respectively. In terms of these eigenvalues,
\begin{equation}
   \mathcal{O}_{d,k} = \frac{1}{4}\sum_{\lambda} e^{-|\lambda\alpha_k(\tau)|^2/2},
\end{equation}
and the total displacement error is,
\begin{subequations}
\begin{gather}
\epsilon_d = \sum_k \epsilon_{d,k},\\
\epsilon_{d,k} = 1 - \left|\frac{1}{4}\sum_{\lambda} e^{-|\lambda\alpha_k(\tau)|^2/2}\right|^2.
\end{gather}
\end{subequations}
If we assume the Lamb-Dicke parameters are equal for each ion and mode involved in the gate ($\eta_{1,k}=\eta_{2,k}=\eta$), then the eigenvalues of $S_{y,k}$ are $\lambda=\{\eta, 0, 0, -\eta\}$, and from the equation above, we find that $\epsilon_{d,k}\approx \eta^2|\alpha_k(\tau)|^2/2$ for small errors.

\subsubsection{Rotation-angle Error}
\label{sec:err_r}
Rotation-angle error $\epsilon_r$ is error in the entangling phase accumulated at the end of the gate. In the phase-space trajectory picture, this can be visualized as the area enclosed by the curves. In previous works, most gate designs do not explicitly target solutions that are robust to rotation-angle error caused by motional frequency drift. In this work, we utilize the contributions from multiple motional modes in order to derive a gate in which the entangling phase is independent
to first order in symmetric detuning offset.

We derive an expression for the rotation error by acting the spin-entangling operator for the gate on the initial state and computing the wavefunction overlap with the target state: $\mathcal{O}_r=\bra{\Phi}\Pi_k e^{i\mathcal{B}_k(\tau)S_{y,k}^2}\ket{00,0}$. Since the spin-entangling operator is independent of the motional state, we only require that the phase accrued during the gate $\theta = \sum_k\eta_{1,k}\eta_{2,k}\mathcal{B}_k(\tau)$ is equal to $\pi/2$ to achieve an MS gate, as shown in equation~(\ref{eq:err_r}). Therefore, any additional phase $\delta\theta=\theta-\pi/2$ results in the rotation $\delta U = e^{i \delta\theta \sigma_{y,1}\sigma_{y,2}/2}$ of the target state, up to an arbitrary global phase, and the rotation error is,
\begin{equation}
    \epsilon_r = 1 - \left|\bra{\Phi}\delta U\ket{\Phi}\right|^2
    = \frac{1}{4}\left|\delta\theta\right|^2. 
\end{equation}

\subsection{Gaussian Pulse Shape}
\label{sec:pulse_shape}

The robustness of our gate design to motional frequency error relies on the specific shape of the pulse amplitude during the gate, which we describe by a time-dependent Rabi rate~$\Omega(t)$. From section~\ref{sec:err_d}, we see that the state infidelity grows linearly with the residual displacement of each mode $|\alpha_k(\tau)|^2$. To reduce these contributions, we note that $\alpha_k(\tau)$ is proportional to the Fourier transform of $\Omega(t)$ evaluated at $\delta_k$, as seen in section~\ref{sec:model}, and we choose to employ a truncated-Gaussian pulse shape,
\begin{equation}
    \Omega(t) = 
\begin{cases}
    \Omega_0 e^{-(t-\tau/2)^2/2z^{2}},& \text{if } 0\le t \le \tau,\\
    0,              & \text{otherwise},
\end{cases}
\end{equation}
where $\Omega_0$ is the peak Rabi rate, and $z$ is the Gaussian width. To understand the effect of this pulse shape, we assume that $z$ is sufficiently small ($z^2\ll\tau^2/8$) such that we can ignore the truncation of $\Omega(t)$ and extend the limits of integration in equation (\ref{eq:alpha}) for $\alpha(\tau)$ to all times,
\begin{subequations}
\begin{gather}
\label{eq:alpha_approx}
\alpha_k(\tau) \approx i\int_{-\infty}^\infty \Omega'(t')e^{-i\delta_k t'}\mathrm{d}t',\\
|\alpha_k(\tau)|^2 \approx 2\pi \Omega_0^2 z^2 e^{-\delta_k^2 z^2},
\end{gather}
\end{subequations}
where $\Omega'(t)=\Omega_0 e^{-(t-\tau/2)^2/2 z^2}$ for all $t$. Here, we see that the Gaussian-like pulse shape guarantees that the displacement error from each mode will exponentially decay with $\delta_k^2$. For the appropriate choice of detuning, this can strongly suppress contributions to $\epsilon_d$ from the modes primarily driving the gate and almost entirely eliminate contributions to $\epsilon_d$ from far-detuned "spectator" modes. 

\subsection{Mode Balancing}

Many MS-gate implementations, including the standard gate with a constant detuning and Rabi rate~\cite{sorensen:2000}, suffer from the displacement error $\epsilon_d$ caused by spectator modes, and the gate performance degrades when the detuning of the mode primarily targeted by the gate has a similar magnitude as the detuning from the other modes. However, since the Gaussian pulse shape strongly suppresses $\epsilon_d$, as discussed in section~\ref{sec:pulse_shape}, we have the freedom to detune close to multiple modes without suffering from large displacement errors. In this section, we take advantage of this freedom and determine a choice of detuning $\delta_c$ that not only maintains the suppression of $\epsilon_d$ but also reduces the rotation-angle error~$\epsilon_r$.

While staying sufficiently far away from all modes ($\delta_k \gg z^{-1}$) to suppress $\epsilon_d$, we aim to balance the contributions to $\theta$ from all modes, such that frequency errors that increase the contribution to $\theta$ from some modes are cancelled to first-order by the decrease in the contribution to $\theta$ from the other modes.
In general, the detunings that balance the contributions to $\theta$ from all modes can be found by solving the following equation for $\delta_\text{c}$,
\begin{equation}
\label{eq:dtheta}
    \frac{\mathrm{d}\theta}{\mathrm{d}\delta_\text{c}}=\sum_k \eta_{1,k}\eta_{2,k}\frac{\mathrm{d}\mathcal{B}_k(\tau)}{\mathrm{d}\delta_\text{c}}=0.
\end{equation}
This equation can be solved numerically to implement gates for which $\epsilon_r$ is first-order robust to frequency error. At least for mono-species ion chains, solutions to equation~(\ref{eq:dtheta}) exist for all pairs of ions.

Although solving equation~(\ref{eq:dtheta}) is straightforward and numerically efficient, we can search for approximate solutions to further simplify the design and implementation of this gate. In the region of $\delta_c$ between two neighboring modes $k_1$ and $k_2$, we can find an approximate solution by neglecting the contribution to $\mathrm{d}\theta/\mathrm{d}\delta_\text{c}$ from all other modes because the $k$-th term in the sum becomes large near $\delta_k=0$ and tends to dominate all other terms. We also find that $\mathrm{d}\mathcal{B}_{k_1}/\mathrm{d}\delta_\text{c}$ and $\mathrm{d}\mathcal{B}_{k_2}/\mathrm{d}\delta_\text{c}$ have the same sign in this region, such that solutions to equation~(\ref{eq:dtheta}) are likely to be found when the products $\eta_{1,k_1}\eta_{2,k_1}$ and $\eta_{1,k_2}\eta_{2,k_2}$ have opposite signs. To the extent that these products have similar magnitudes, the two modes make similar contributions to the gate when $|\delta_{k_1}|\approx|\delta_{k_2}|$, and we can expect to find a solution to equation~(\ref{eq:dtheta}) that is close to the midpoint between modes $k_1$ and $k_2$: $\delta_\text{c}=(\nu_1 + \nu_2)/2$.
In practice, one can use this value of $\delta_c$ as a starting point and experimentally calibrate $\delta_c$ as we describe in section~\ref{sec:ExpResults}.


\subsection{Gate Comparison}
\label{sec:gate_compare}

\begin{figure*}[t]
    \centering
    \raggedright
    \begin{subfigure}[b]{0.329\textwidth}
        \raggedright
        \resizebox{1.0\textwidth}{!}{
        \includegraphics{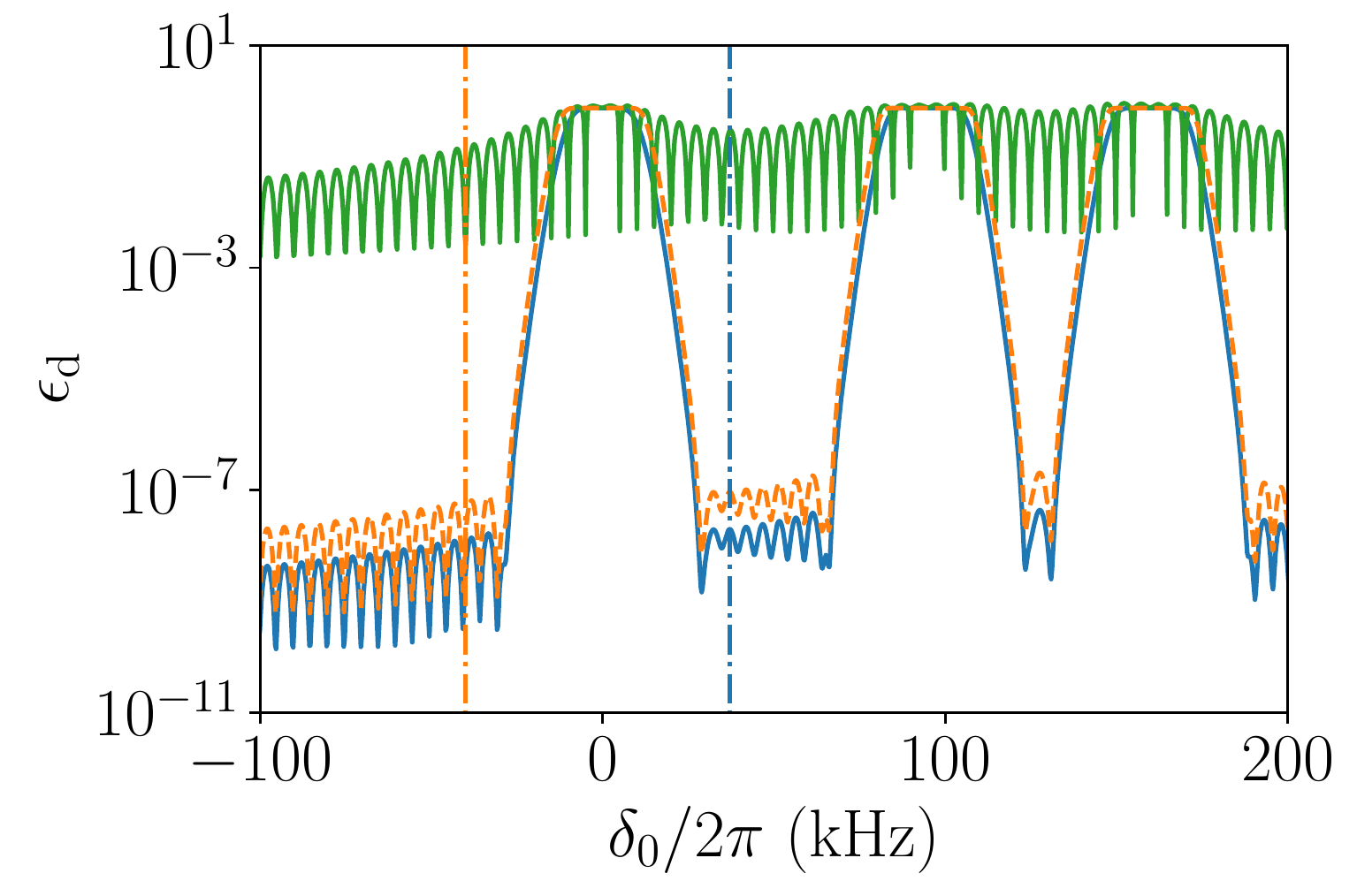}}
        \vspace{-3em}
        \caption{\raggedright}
    \end{subfigure}
    \begin{subfigure}[b]{0.329\textwidth}
        \raggedright
        \resizebox{1.0\textwidth}{!}{
        \includegraphics{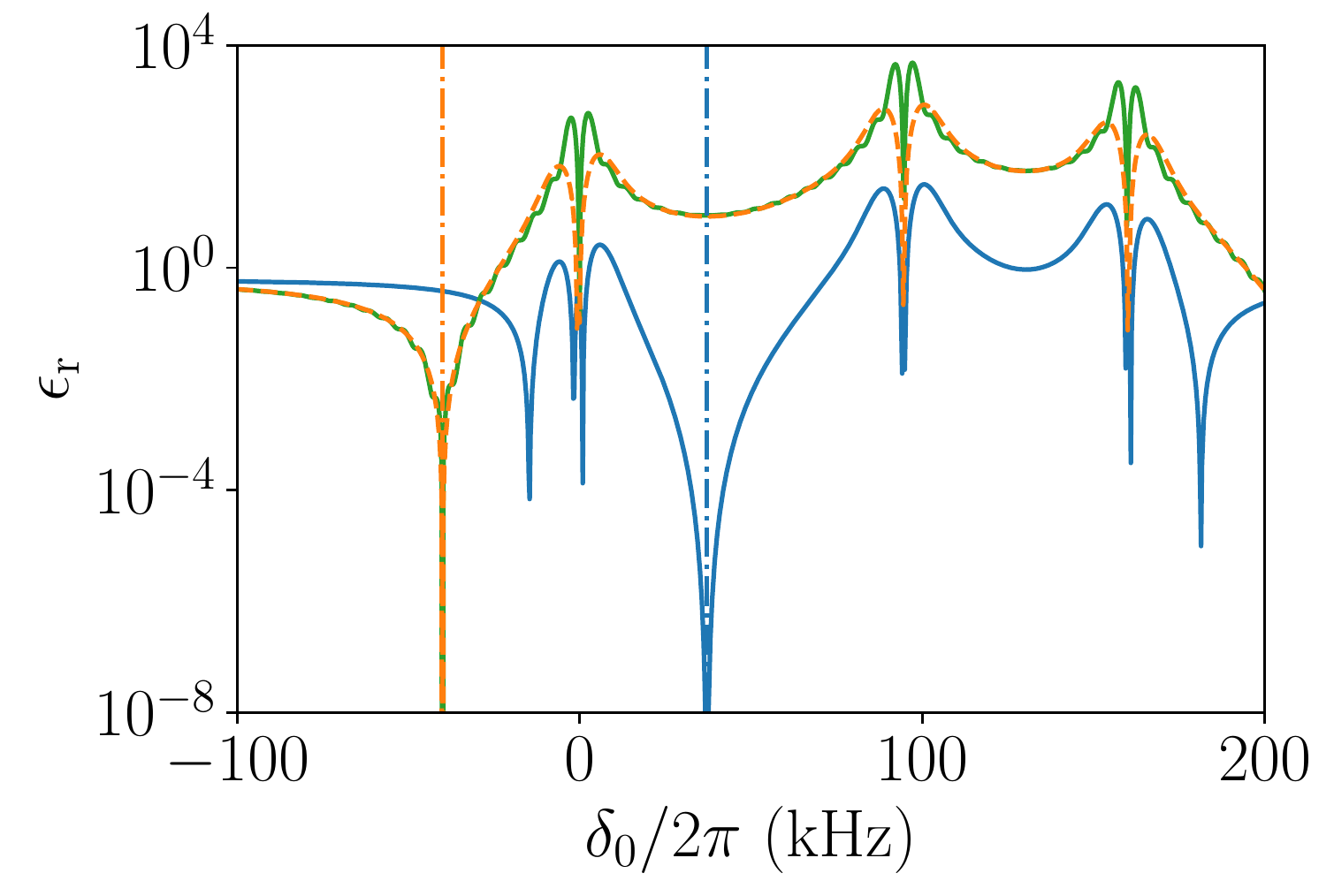}}
        \vspace{-3em}
        \caption{\raggedright}
    \end{subfigure}
    \begin{subfigure}[b]{0.329\textwidth}
        \raggedright
        \resizebox{1.0\textwidth}{!}{
        \includegraphics{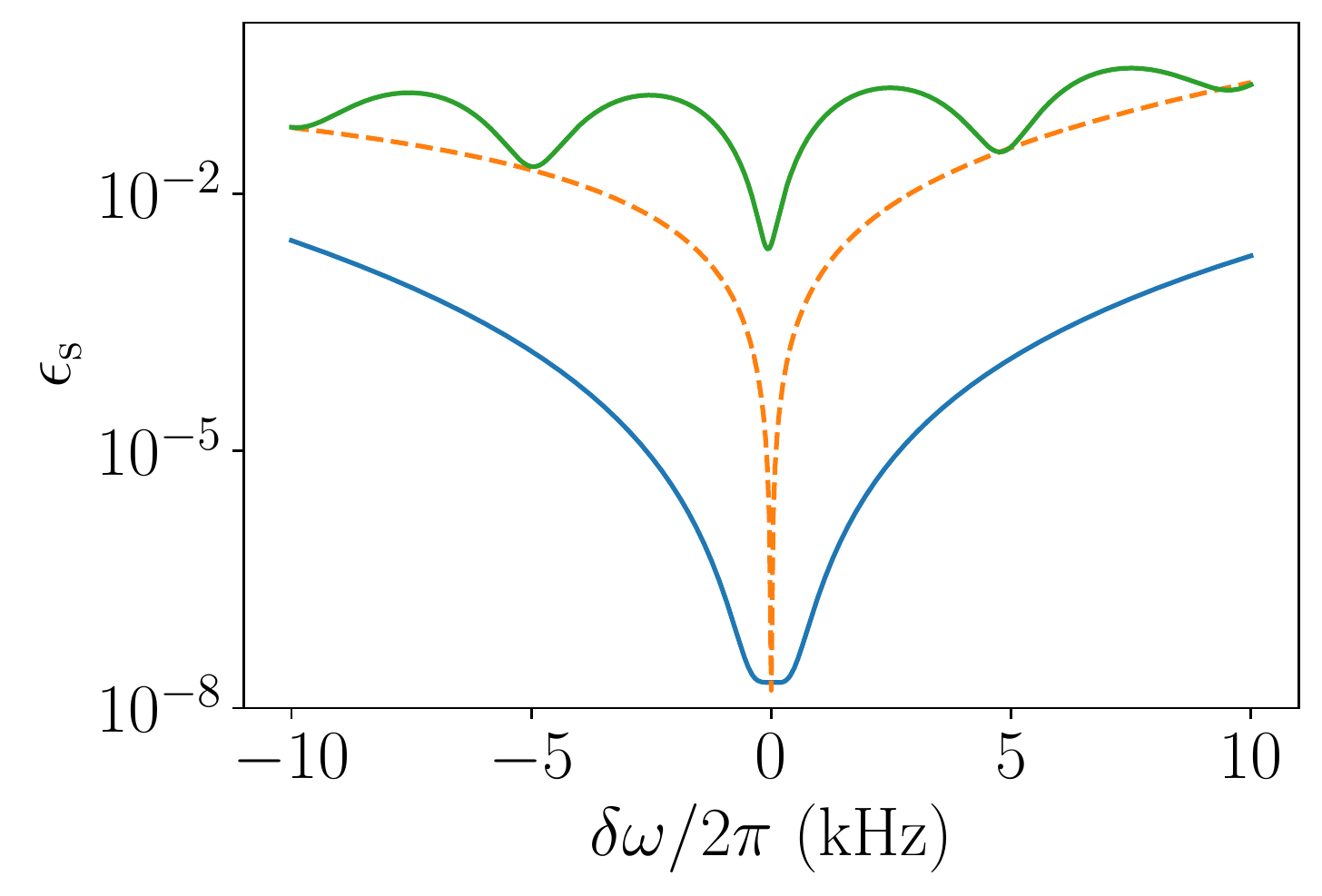}}
        \vspace{-3em}
        \caption{\raggedright}
    \end{subfigure}
    \caption{\raggedright (a) Displacement error $\epsilon_d$, (b) rotation-angle error $\epsilon_r$, and (c) state infidelity~$\epsilon_s=\epsilon_d+\epsilon_r$ for a three-ion chain with three different pulse shapes. From bottom to top at most detunings, the curves correspond to (solid blue) a Gaussian pulse with a symmetric detuning of $\delta_0/2\pi=37.2$~kHz, (dashed orange) a Gaussian pulse with $\delta_0/2\pi=-40$~kHz, and (solid green) a square pulse with $\delta_0/2\pi=-40$~kHz. The detuning $\delta_0=0$ is resonant with the lowest frequency motional mode, and the next two higher motional modes are at $\delta_0/2\pi = 94.7$~kHz and $\delta_0/2\pi = 160$~kHz, corresponding to the peaks in both error metrics. Vertical dash-dotted lines are drawn at (orange) $\delta_0/2\pi=-40$~kHz and (blue) $\delta_0/2\pi=37.2$~kHz. The frequency error $\delta\omega/2\pi$ is defined relative to the optimal value of $\delta_0/2\pi$ for each pulse.}
    \label{fig:displacement_and_angle}
\end{figure*}

To demonstrate the robustness to frequency error of the balanced Gaussian gate,
we simulate MS gates with three different pulse shapes subject to the symmetric frequency error~$\delta\omega$ and compute the contributions to the state infidelity from displacement error~$\epsilon_d$ and rotation-angle error~$\epsilon_r$ for each pulse. We compare the balanced Gaussian pulse, which has a detuning that solves equation ~(\ref{eq:dtheta}), with a standard Gaussian pulse and a square pulse (constant Rabi rate during the gate) that are both detuned below the lowest motional mode. The duration of each gate is 200~$\mu$s and the width of each Gaussian pulse is $z=25$~$\mu$s. Being a small fraction of the pulse length, this choice of $z$ creates Gaussian-like pulses with small truncation effects.

We provide a concrete example for this comparison by simulating each gate on the outer ions of a three-ion chain with an ion separation of $4.5$~$\mu$m and radial trapping frequencies of $\omega_\text{trap, ra}/2\pi=2.52$~MHz and $\omega_\text{trap,rb}/2\pi=2.19$~MHz in the orthogonal radial-a and radial-b directions, respectively. These trapping frequencies correspond to the motional frequency of the center-of-mass (highest) mode in each direction and have values that are typical for the surface trap of the QSCOUT testbed. We also assume the laser $k$-vector is aligned at a 45$\degree$ angle between the two radial directions, and we neglect any excitation of the axial motional modes, which will have a large detuning compared to the radial modes during the gate and are orthogonal to the $k$-vector of the laser. 

We design the balanced Gaussian gate by targeting the lowest two radial modes: the zig-zag ($k=0$) and tilt ($k=1$) radial-b modes, which have a splitting of $\Delta \nu_{10}/2\pi=(\nu_1-\nu_0)/2\pi=94.7$~kHz. The targeted ions have Lamb-Dicke parameters that are equal in the zig-zag mode, while being equal and opposite in the tilt mode, such that the products $\eta_{1,0}\eta_{2,0}$ and $\eta_{1,1}\eta_{2,1}$ have opposite signs and are approximately equal in magnitude. We then solve equation~(\ref{eq:dtheta}) in the region between these modes and obtain a detuning of $\delta_0/2\pi=37.2$~kHz above the lowest frequency mode (zig-zag). For the standard Gaussian and square gates, we choose a detuning of~$\delta_0/2\pi=-40$~kHz. As a final step, we select the peak Rabi rate $\Omega_0$ for each gate separately to guarantee $\theta=\pi/2$ at the chosen detuning. The frequency error~$\delta\omega$ is defined relative to these nominal detunings for each gate. 

Figure~\ref{fig:displacement_and_angle} shows the simulated values of $\epsilon_d$ and $\epsilon_r$ for the three different pulse shapes as a function of the detuning~$\delta_0/2\pi$ from the lowest motional mode. The Gaussian pulse shape (dashed orange and solid blue) suppresses displacement error as long as the gate is performed sufficiently far from all motional modes. By contrast, the square pulse gate (solid green) has narrow minima, and relatively high displacement error persists at all detunings in the range displayed. A broad dip in the rotation-angle error is apparent for the balanced Gaussian gate (solid blue) when the derivative with respect to detuning of the sum of contributions of multiple modes goes to zero. Intuitively, this can be described by the fact that as the detuning moves in one direction away from the optimal point the contribution from one mode becomes smaller but the contribution from the other mode becomes larger. By contrast, the standard Gaussian gate 
(dashed orange) and the square gate (solid green) only have narrow dips in rotation-angle error and thus have a small detuning range over which the target phase is accumulated. For small frequency errors, the state infidelity $\epsilon_s=\epsilon_d+\epsilon_r$ for each Gaussian pulse is strongly dominated by the contribution from the rotation-error $\epsilon_r$. Figure~\ref{fig:displacement_and_angle}c shows $\epsilon_s$ for each pulse shape over a range of experimentally relevant frequency errors, demonstrating the improved robustness to $\delta\omega$ when using the balanced Gaussian pulse shape.


\section{Gate Implementation}

\subsection{Gate Parameter Selection}
\label{sec:gate_params}

In addition to choosing the correct detuning to balance the contributions of multiple motional modes, the time-domain standard deviation, $z$, of the truncated Gaussian pulse shape is a free parameter that may be tuned to optimize the gate performance. This may be done numerically or empirically given sufficient intuition about the contributions to gate error. In particular, we find that optimal value of $z$ for robustness to frequency error depends on the detuning from the closest two motional modes. On the one hand, the truncation of the Gaussian lying outside of the pulse duration $\tau$ leads to some amount of square pulse character with an abrupt pulse turn on. This truncation effect and square pulse character lead to some residual displacement error, especially when the Gaussian width becomes comparable to the gate time ($z/\tau \approx \mathcal{O}(1)$). This error can be seen as the oscillating floor in the plot of $\epsilon_d$ in Fig.~\ref{fig:displacement_and_angle}a. On the other hand, displacement error can become large when $z$ is small enough that $e^{-\delta^2_{k}z^2}$ becomes significant. This can be understood as Fourier broadening of the pulse as it becomes narrow in time. To optimize robustness, a value of $z$ can be chosen that balances the infidelity contribution from cutoff effects and Fourier broadening over a given range of frequency error. One must also take into consideration that as $z$ is reduced, the peak intensity of the pulse can become an experimental challenge because the Rabi rate must be scaled up (by increasing $\Omega_0$) to achieve the proper rotation angle of $|\theta| = \pi/2$.

Additionally, the choice of modes targeted by the gate plays an important role in its robustness to frequency error. In general, motional modes with a larger frequency splitting~$\Delta \nu/2\pi$ provide more robustness by allowing larger detunings ($\delta_{k_1}$ and $\delta_{k_2}$) through which the displacement error $\epsilon_d$ becomes more strongly suppressed. Also, larger detunings reduce the magnitude of $\mathrm{d}\theta/\mathrm{d}\delta_c$ away from its zero crossing, providing a reduction in the rotation-angle error $\epsilon_r$ for finite frequency errors. For typical ion chains in a harmonic well, the lowest two radial modes will have the largest splitting and therefore the maximum robustness to frequency error. This situation is reversed in the case of axial modes, for which the highest two modes typically have the largest splitting. 

\begin{figure}[ht]
    \raggedright
    \begin{subfigure}[b]{0.47\textwidth}
    \raggedright
    \resizebox{1.0\textwidth}{!}{
    \includegraphics{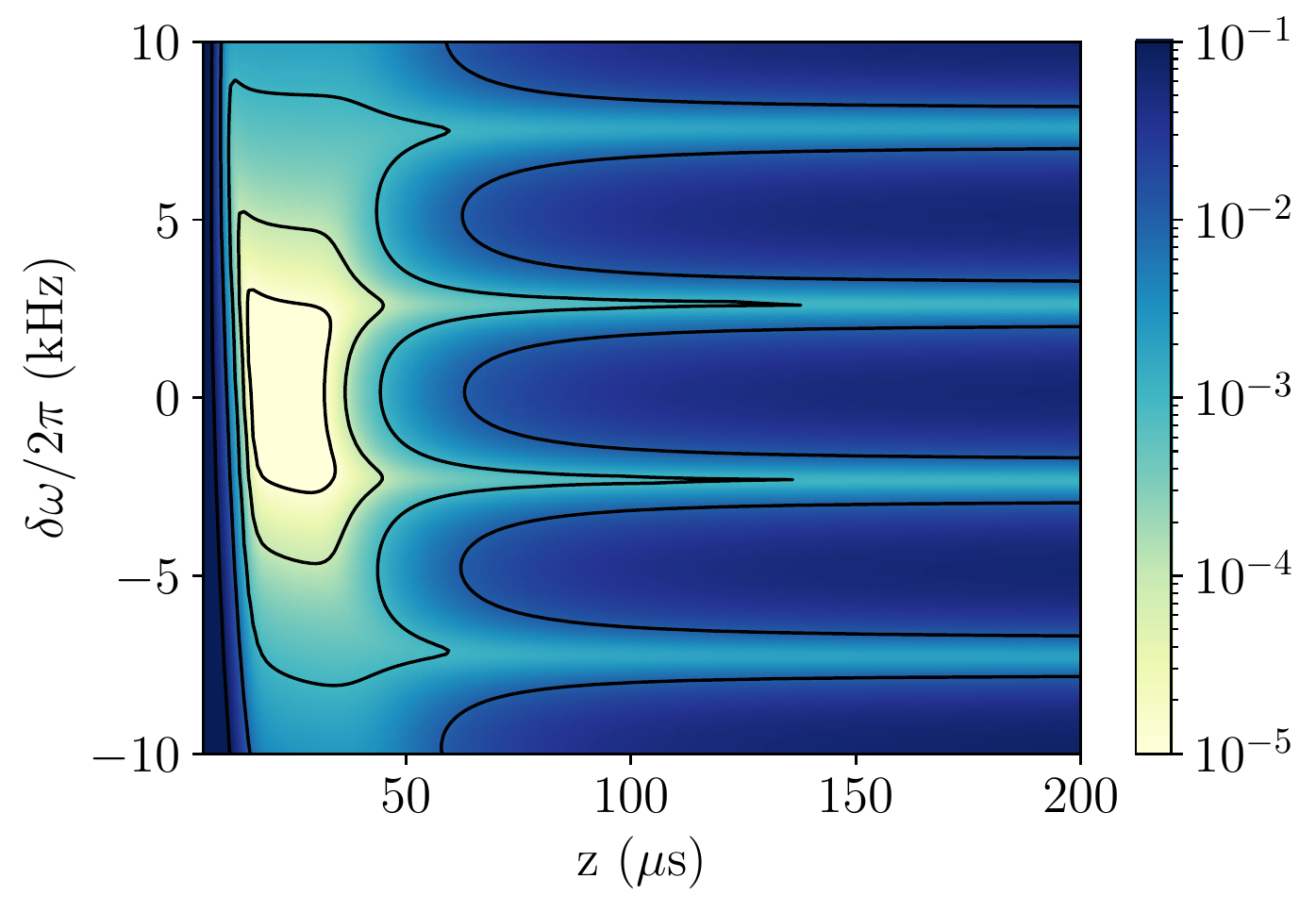}}
    \end{subfigure}
    \caption{\raggedright State infidelity $\epsilon_s$ as a function of the Gaussian width $z$ and symmetric detuning offset $\delta\omega$ for a truncated, balanced Gaussian MS gate on a three-ion chain. As the gate duration is $\tau=200$\,$\mu$s, the gate varies from delta-function to square-pulse character from left to right.
    }
    \label{fig:contour}
\end{figure}

Because the gate parameters $\delta_c$, $z$, and $\Omega_0$ depend on the specific distribution of motional modes, we provide a concrete example of optimal parameter selection for the model of a three-ion chain described in section~\ref{sec:gate_compare}, which targets the outer ions and the lowest two radial modes. To determine the optimal gate parameters and assess their robustness for this model, we numerically vary the Gaussian width $z$ for a fixed pulse duration of $\tau = 200\,\mu$s and show the performance of the gate over a range of $\delta\omega = \pm 10\,$kHz in Fig.~\ref{fig:contour}. At large values of $z$ in this plot, the pulse resembles a square pulse with an approximately constant magnitude during the gate and a strong truncation effect. In this regime, we recover an infidelity proportional to ${\rm sinc}(\delta\omega \tau)$, as we expect for the Fourier transform of a square pulse. Toward the lower end of $z$ on the plot, the pulse resembles a Gaussian with a small truncation effect, and we find that there is a broad region of $z$ and $\delta\omega$ where the gate performs well. For example, the state infidelity $\epsilon_s$ remains below $10^{-3}$ over the range: $-7.8$\,kHz~$\le \delta_{\omega}/2\pi\le 8.5$\,kHz for $z=25$\,$\mu$s and over the range: $13$\,$\mu$s~$\le z \le 44$\,$\mu$s for $\delta\omega=0$.


\begin{figure*}[ht]
    \centering
    \raggedright
    \begin{subfigure}[b]{0.32\textwidth}
        \raggedright
        \resizebox{1.0\textwidth}{!}{
        \includegraphics{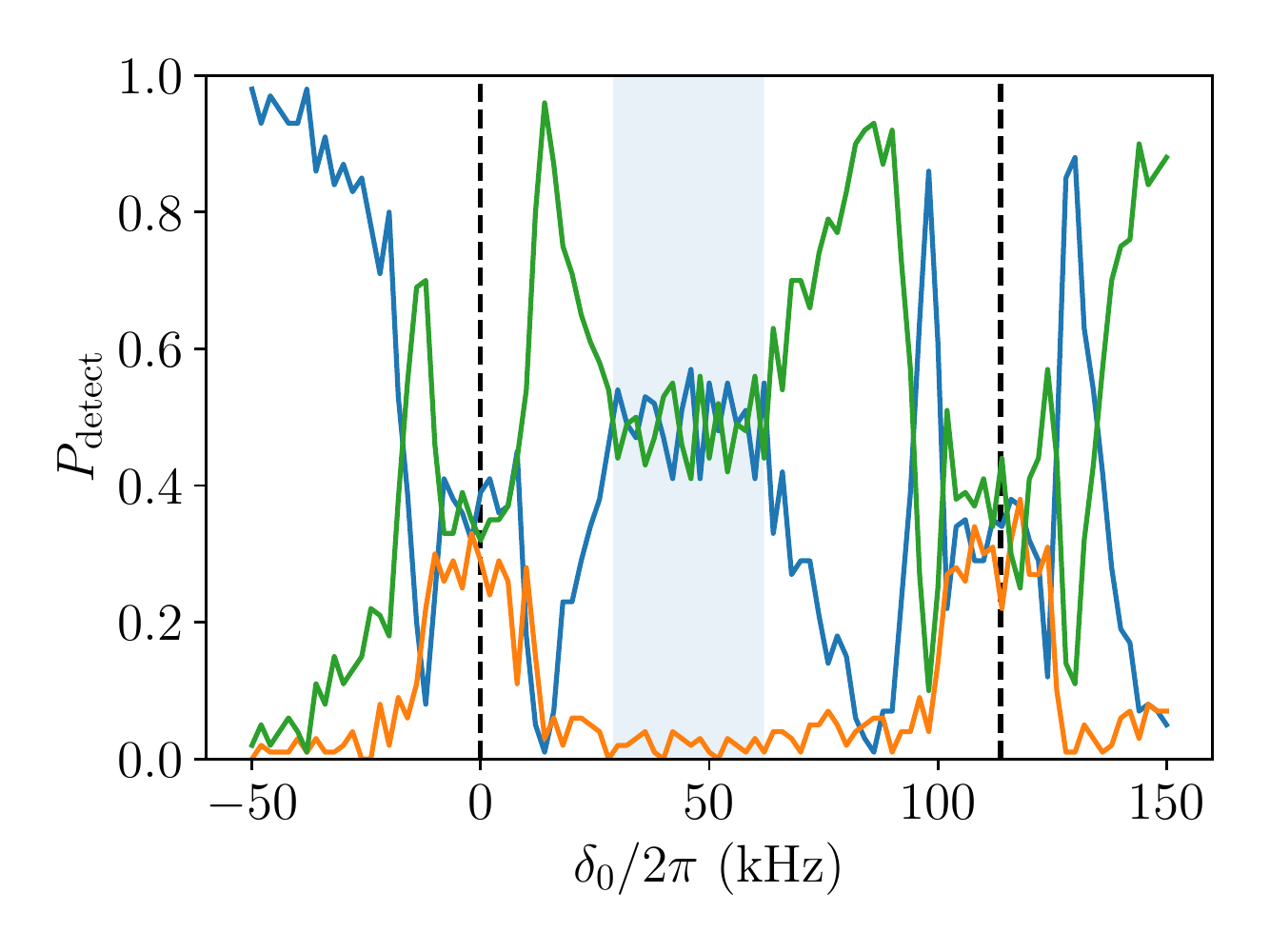}}
        \vspace{-3em}
        \caption{\raggedright}
    \end{subfigure}
    \begin{subfigure}[b]{0.32\textwidth}
        \raggedright
        \resizebox{1.0\textwidth}{!}{
        \includegraphics{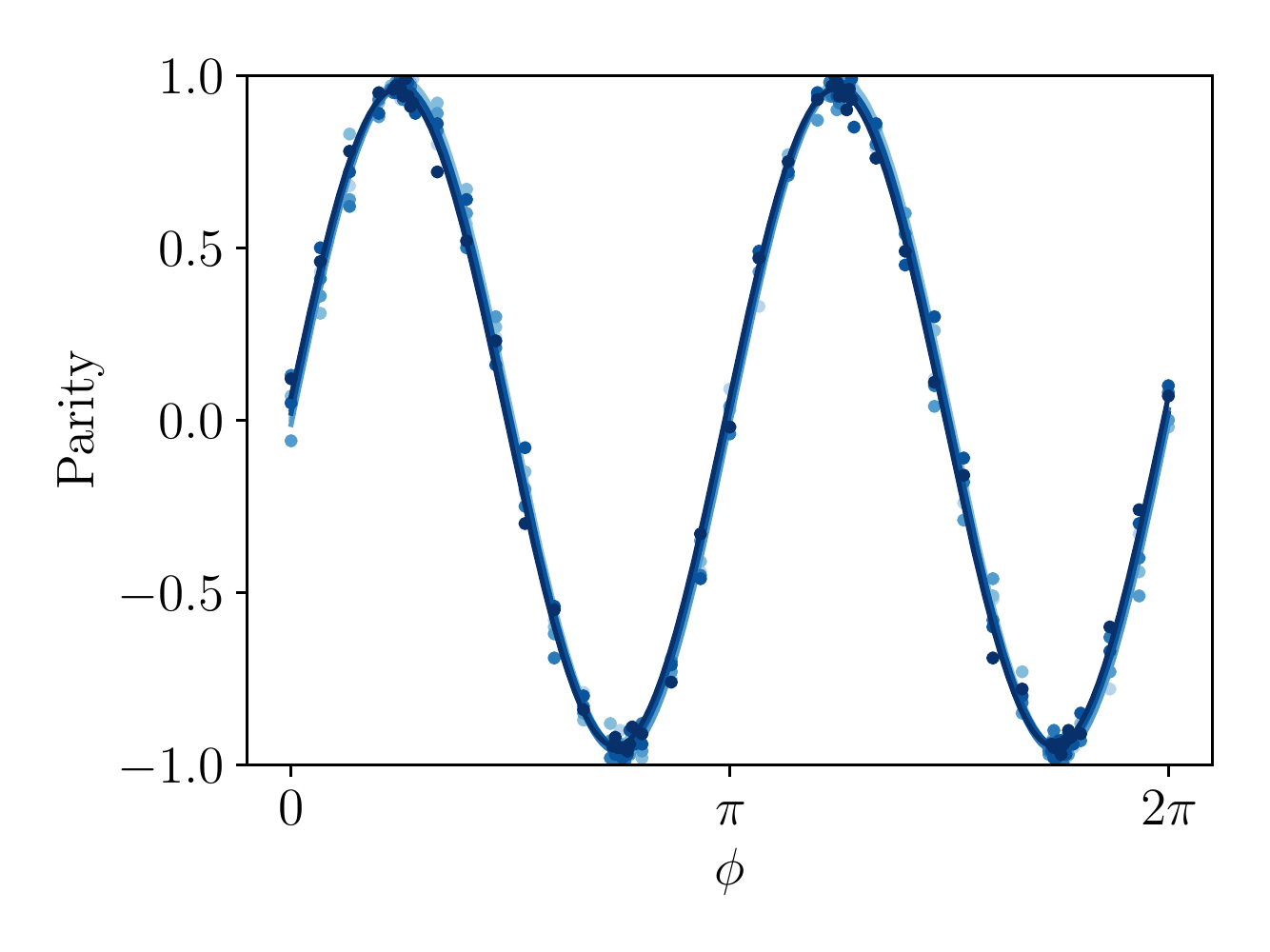}}
        \vspace{-3em}
        \caption{\raggedright}
    \end{subfigure}
    \begin{subfigure}[b]{0.32\textwidth}
        \raggedright
        \resizebox{1.0\textwidth}{!}
        {
        \includegraphics{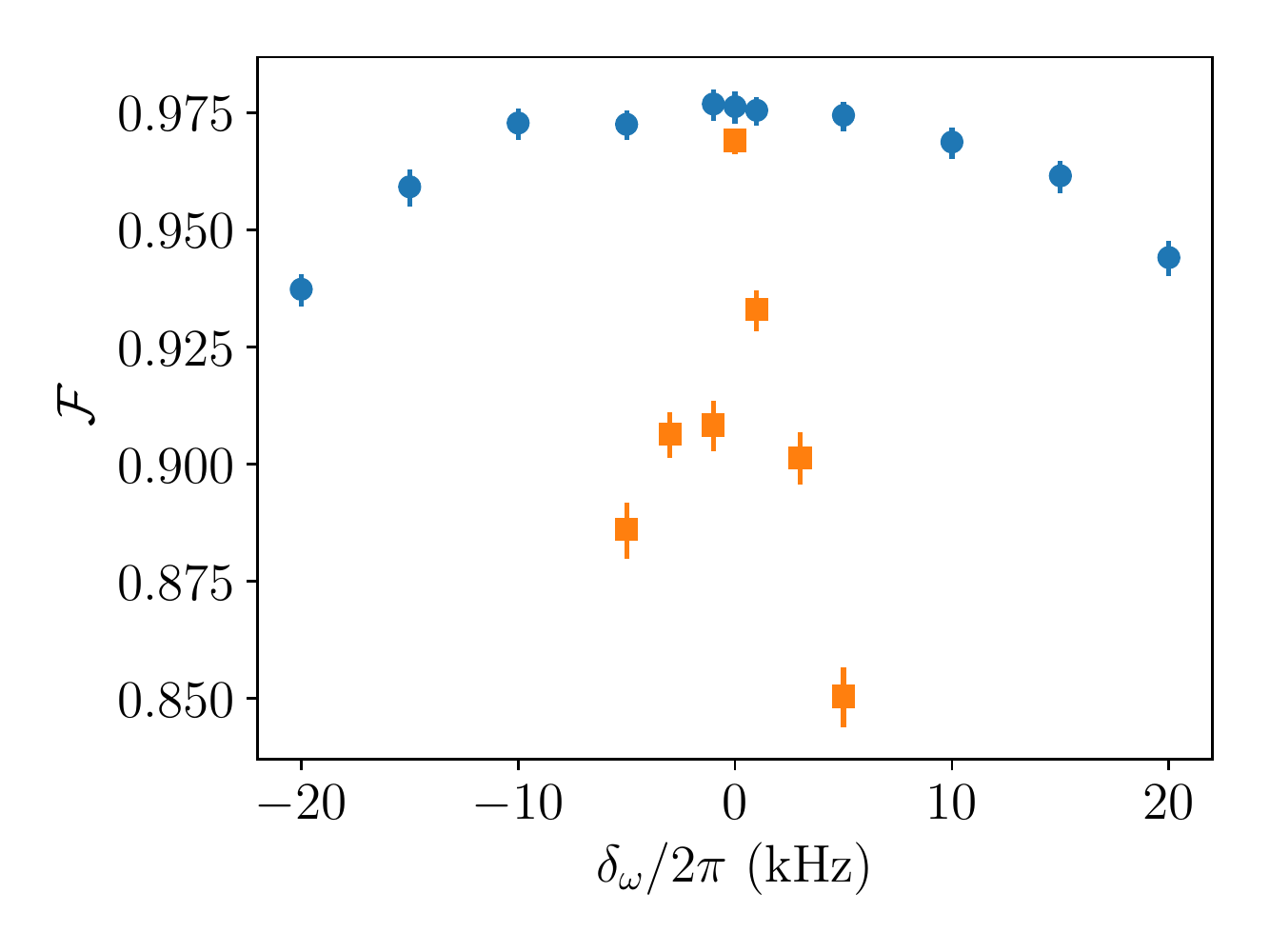}
        }
        \vspace{-3em}
        \caption{\raggedright}
    \end{subfigure}
    \caption{\raggedright 
    (a) A symmetric detuning scan between two motional modes (marked with dashed lines at 0\,kHz and $\approx$110\,kHz) shows a broad region around $\delta_{0}/2\pi = 46$\,kHz where the balanced Gaussian gate performs well (shaded region). Lines top to bottom at $\delta_{0}/2\pi < -25\,$kHz are the detection probabilities $P(00)$, $P(11)$, and $P(01)+P(10)$ after preparing in $\ket{00}$ and applying the gate. 
    (b) Parity oscillations are measured (points) by applying a single-qubit rotation with variable phase ($\phi$) and fit (lines) to a sine. From light to dark, the symmetric detuning offset increases over the range $\delta_{\omega}/2\pi=\pm$10\,kHz for the balanced Gaussian gate.
    (c) Measured entangling gate 
    fidelity ($\mathcal{F}$)
    as a function of symmetric detuning offset ($\delta_{\omega}$) for the balanced Gaussian (blue circles) and unbalanced Gaussian (orange squares). Fidelity is calculated according to Eq.~\ref{eq:ExperimentFidelity} using the parity scan results and population measurements. Uncertainty is derived from a Wilson score interval on the population measurement and fitting uncertainty on the parity scan.
    }
    \label{fig:experiment_terr_vs_delta}
\end{figure*}

\subsection{Experimental Implementation}

We implement the derived gate on a chain of ${N=3}$ 171Yb$^{+}$ ions to measure robustness to symmetric detuning offsets. We use the hyperfine ground, ``clock'' states as the qubit levels: $\ket{F=0, m_{F}=0} \equiv \ket{0}$ and $\ket{F=1, m_{F}=0} \equiv \ket{1}$. In all experiments, all ions are initialized in $\ket{0}$ and the gate under study is applied to two target ions. 
A global, single-qubit $\pi/2$ rotation is then applied in parity scan measurements (described in \ref{sec:ExpResults}). Finally, in all measurements, the population of each qubit state is determined by fluorescence detection. The relevant two-ion state is labelled as ${\rm Tr}_{c}(\ket{a}\otimes\ket{b}\otimes\ket{c})\equiv\ket{ab}$ where $a$ and $b$ are the states of the two target ions, and $c$ is the state of the third ``spectator'' ion in the chain, which is ignored.

We set the principal axes of the trap to be at a 45$\degree$ angle from the effective Raman $k$-vector to allow for Raman sideband cooling on all radial modes, and thus observe a total of $2N$ radial motional frequencies. For the fidelity measurements presented here, we have radial motional frequencies of $\nu_\text{ra}/2\pi=\{2.134, 2.229, 2.296\}$\,MHz and $\nu_\text{rb}/2\pi = \{1.832, 1.941, 2.017\}$\,MHz, and an axial center-of-mass frequency of $
\nu_\text{axial}/2\pi =$ 0.52\,MHz. As modelled previously, we operate between the lowest two radial-b modes. 

The Gaussian pulse shape is approximated by a natural cubic spline with 13 amplitude knots that are passed to our custom Radio Frequency System-on-Chip (RFSoC) hardware, ``Octet'' \cite{QSCOUTManual}. Octet generates the RF waveform that drives the acousto-optic modulators (AOMs) which perform the required RF to optical transduction. The spline knots are equally spaced along the square root of a Gaussian pulse shape and applied to both the individual addressing beams and counterpropagating global beam AOMs, thus producing a Gaussian temporal profile in the two-photon Raman Rabi rate of the pulse. The first and last knots of the spline are non-zero, and truncate parts of the infinite Gaussian lying outside of the pulse duration. 
 
We choose to implement a gate with $z/\tau=26.5\,\mu\text{s}/200\,\mu\text{s} \approx 0.13$ on the experiment to maintain reasonably low truncation error while still avoiding demanding Rabi rates. A gate time of 200\,$\mu$s is chosen empirically to minimize the effects of heating, truncation, and Fourier broadening. 
The ``balance point'' detuning ($\delta_0$) is found by scanning the symmetric detuning and finding the point of zero slope in the $\ket{11}$ population, and the peak Rabi rate ($\Omega_{0}$) is set by varying the scaling of the Gaussian pulse shape and finding the point of equal $\ket{00}$ and $\ket{11}$ populations. 
For the balanced Gaussian gate demonstrated here, $\delta_{0}/2\pi = 46\,$kHz and $\Omega_{0} \approx 2\pi \times 190\,$kHz. 
 
Although the wavelength (355\,nm) of the Raman laser is chosen to approximately balance the AC Stark Shift on the qubit transition, there remains a residual differential AC Stark Shift on the order of 1\,kHz. To compensate for this shift, we dynamically apply a virtual frame rotation at a rate proportional to the intensity of the laser applied to each ion. The magnitude of the frame rotation is found by preparing in the $\ket{00}$ state, applying two MS gates back to back, and maximizing the resultant population in the $\ket{11}$ state. 

\subsection{Experimental Results}
\label{sec:ExpResults}
We measure the performance of the balanced Gaussian MS gate and compare it to the unbalanced Gaussian MS gates. While our heating rate is high and ultimate fidelity of both gates appears to be dominated by incoherent heating errors, there is clear difference in response to intentionally applied symmetric detuning offsets. At several points in the measured range of detuning offsets, we measure both the even parity population after an MS gate and the amplitude of a parity oscillation acquired by applying a single-qubit $\pi/2$ pulse with variable phase after the MS gate. We then estimate the fidelity ($\mathcal{F}$) according to \cite{Sackett2000, Kim2009, ManningThesis, Figgatt2019},
\begin{equation}
   \mathcal{F} = \frac{1}{2}(\rho_{00}+\rho_{11}) + \frac{1}{2}A_{\pi} ,
   \label{eq:ExperimentFidelity}
\end{equation}
where $\rho_{ab}$ is the population of the $\ket{ab}$ state after the MS gate and $A_{\pi}$ is the amplitude of the parity oscillation.

We apply the gate to the outer two ions on a three-ion chain, operating between the zig-zag and tilt modes and characterize its performance, as presented in Fig.~\ref{fig:experiment_terr_vs_delta}. In a scan of the symmetric detuning offset, we prepare the ions in $\ket{00}$, the gate is applied, and the populations are read out. The even parity populations serve as an indicator for gate angle, and show the correct gate angle is achieved in a range of $\approx$ 20\,kHz centered around $\delta_{0}/2\pi = 46$\,kHz 
relative to the zig-zag mode, when ${\rho_{00} \approx 0.5 \approx \rho_{11}}$. The odd parity population ($\rho_{01}+\rho_{10}$) is an indicator of displacement error, and shows good performance (population is near zero) as long as the detuning is sufficiently far from the motional modes. We then estimate the fidelity of the gate at various symmetric detuning offsets by taking a parity curve and a population measurement with 4000 shots at each detuning. We find that the fidelity of the balanced Gaussian MS gate drops by < 1\% over $\delta_{\omega}/2\pi \leq \pm 10\,$kHz, indicating broad robustness to symmetric detuning offset. By contrast, the fidelity of an unbalanced Gaussian gate (operated below the lowest mode, $\delta_{0}/2\pi = -25.3\,$kHz \footnote{The laser power required to perform the unbalanced Gaussian MS gate is significantly higher since there is only strong contribution from one motional mode instead of two. As such, we are limited to a detuning of -25.3\,kHz to stay within the bounds of our laser system. We also use the center ion and an edge ion in the unbalanced case to take advantage of the stronger zig-zag coupling from the center ion.}) drops by > 3\% even from an offset of $\pm 1$\,kHz. While the peak fidelity of the balanced Gaussian is only 97.7\raisebox{0.5ex}{\tiny$^{+0.3}_{-0.4}$}\%, we currently have relatively high heating rates (1000s of quanta/s) compared to contemporary systems \cite{EganThesis, BrownMaterials2021}, and we believe that our peak fidelity is limited mostly by incoherent heating errors. 

\subsection{Extension to Larger Numbers of Ions}

In addition to numerically and experimentally verifying performance of the balanced Gaussian gate on a chain of three ions, we also numerically explore the simulated MS gate performance on chains of variable length from $N=2$ up to $N=33$. As described in section~\ref{sec:gate_params}, we target the lowest two motional modes ($k=0$ and $k=1$) of the $N$-ion chain because we expect to achieve more robustness to frequency error $\delta\omega$ by targeting the neighboring modes with the largest frequency splitting. In the region between these modes, we will find a detuning $\delta_c$ that solves equation~(\ref{eq:dtheta}) and balances the contributions to $\epsilon_r$ from all modes, which is possible when the products $\eta_{1,0}\eta_{2,0}$ and $\eta_{1,1}\eta_{2,1}$ have opposite signs. 

To maintain consistency in the sign and magnitude of the Lamb-Dicke parameters between chains of different $N$, we target the ions to the immediate left and right of center, for which we find a solution to equation~(\ref{eq:dtheta}) in each case. The targeted ions are the two center ions of the chain for even $N$ and the center ion and one of its neighbors for odd $N$. For this choice of ions, targeting the lowest two modes leads to a balanced gate with a value of $\delta_c$ that solves equation~(\ref{eq:dtheta}); however, $\theta$ has the opposite sign for even $N$ than for odd $N$. To maintain the same value of the target propagator $U(\tau)=e^{-i \sigma_{y,1}\sigma_{y,2}\theta/2}$ (and hence the same target gate) in our simulations for both even and odd $N$, we impose a differential laser phase of $\pi$ between the two ions for even $N$. We model this change in phase by $\sigma_{y,2}\rightarrow-\sigma_{y,2}$, which modifies the handedness of the phase-space trajectories and has the same effect on $U(\tau)$ as $\theta\rightarrow-\theta$.


As $N$ grows, we relax the axial frequency of the harmonic potential such that at all $N$, the separation between the equilibrium positions of the two center ions (the center ion and its neighbors) for even (odd) $N$ is a fixed value. We call this quantity the center ion separation parameter, $\Delta x_0$. We make this choice to keep a fixed center ion separation parameter at the expense of varying axial frequency in order to maintain consistency with an experimental apparatus designed for a fixed individual addressing beam separation.
This choice comes at the cost of an increasingly dense radial motional mode sideband spectrum as $N$ grows because the spacing of the radial modes is set by the ratio of the axial to radial trap frequencies. 
We also note that at any particular $N$, the separation of ions will increase from center to edge of the chain, as we model a simple harmonic potential in the axial direction. While anharmonic potentials have been considered in other works to keep spacing constant within a chain \cite{Lin2009, Johanning2016, Xie2017}, we deem the implementation of a non-harmonic potential to be outside the scope of this investigation. 
We also use the same values of $\tau=200$\,$\mu$s and $z=25$\,$\mu$s for each $N$ and $\Delta x_0$.

\begin{figure}
    \raggedright
    \begin{subfigure}[b]{0.235\textwidth}
    \raggedright
    \resizebox{1.0\textwidth}{!}{
    \includegraphics{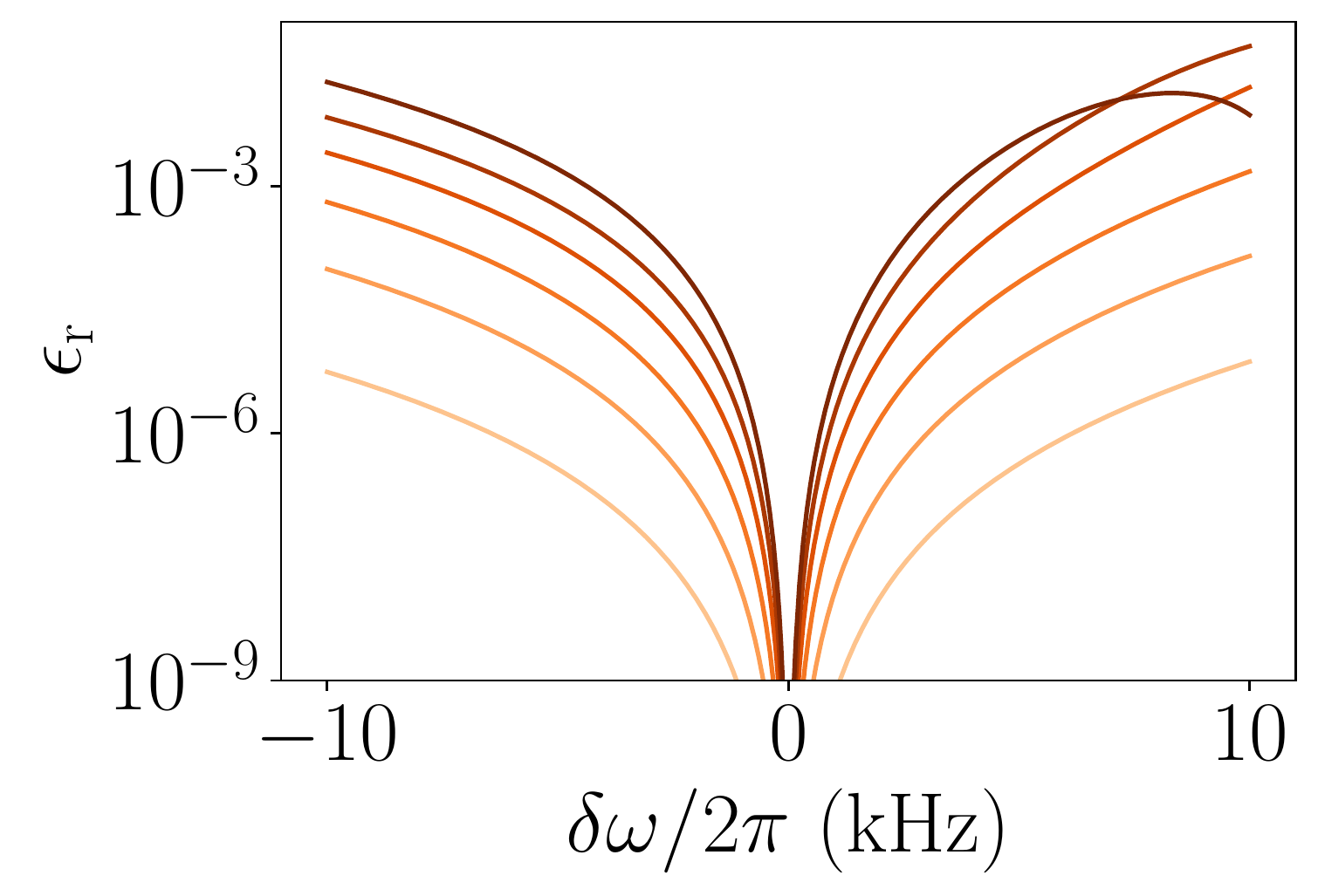}}
    \vspace{-3em}
    \caption{\raggedright}
    \end{subfigure}
    \begin{subfigure}[b]{0.235\textwidth}
    \raggedright
    \resizebox{1.0\textwidth}{!}{
    \includegraphics{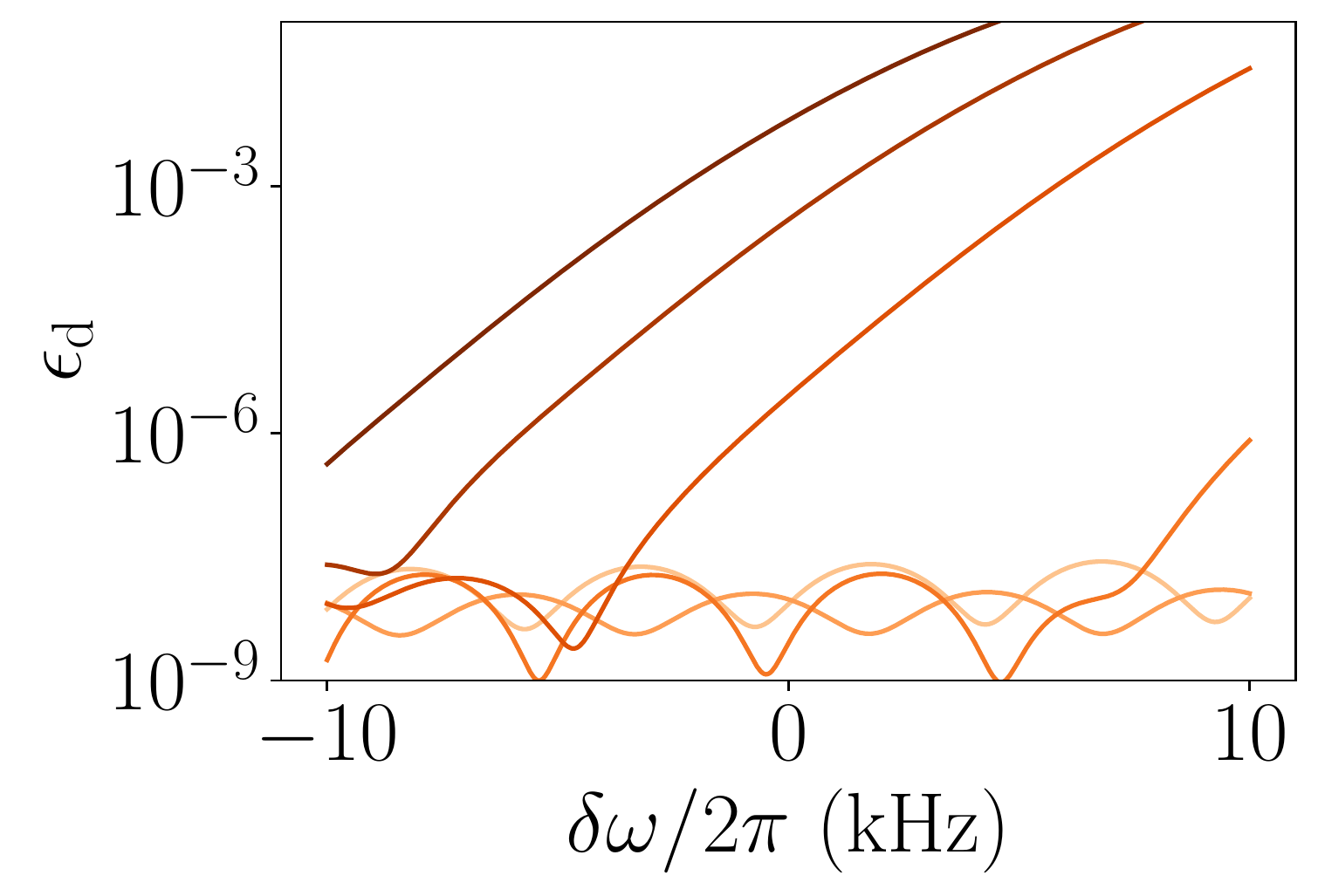}}
    \vspace{-3em}
    \caption{\raggedright}
    \end{subfigure}
        \hfill
    \begin{subfigure}[b]{0.235\textwidth}
    \raggedright
    \resizebox{1.0\textwidth}{!}{
    \includegraphics{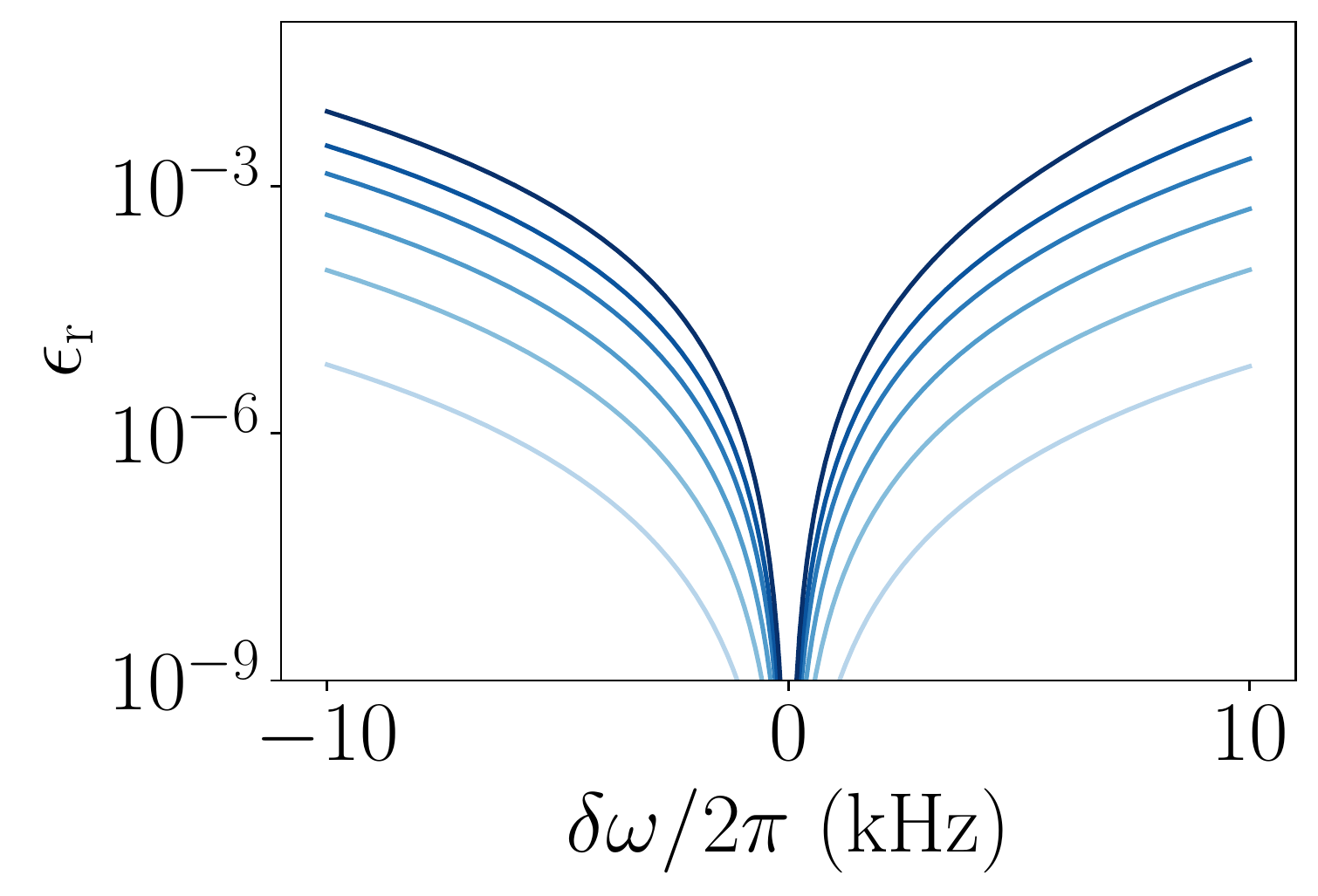}}
    \vspace{-3em}
    \caption{\raggedright}
    \end{subfigure}
    \begin{subfigure}[b]{0.235\textwidth}
    \raggedright
    \resizebox{1.0\textwidth}{!}{
    \includegraphics{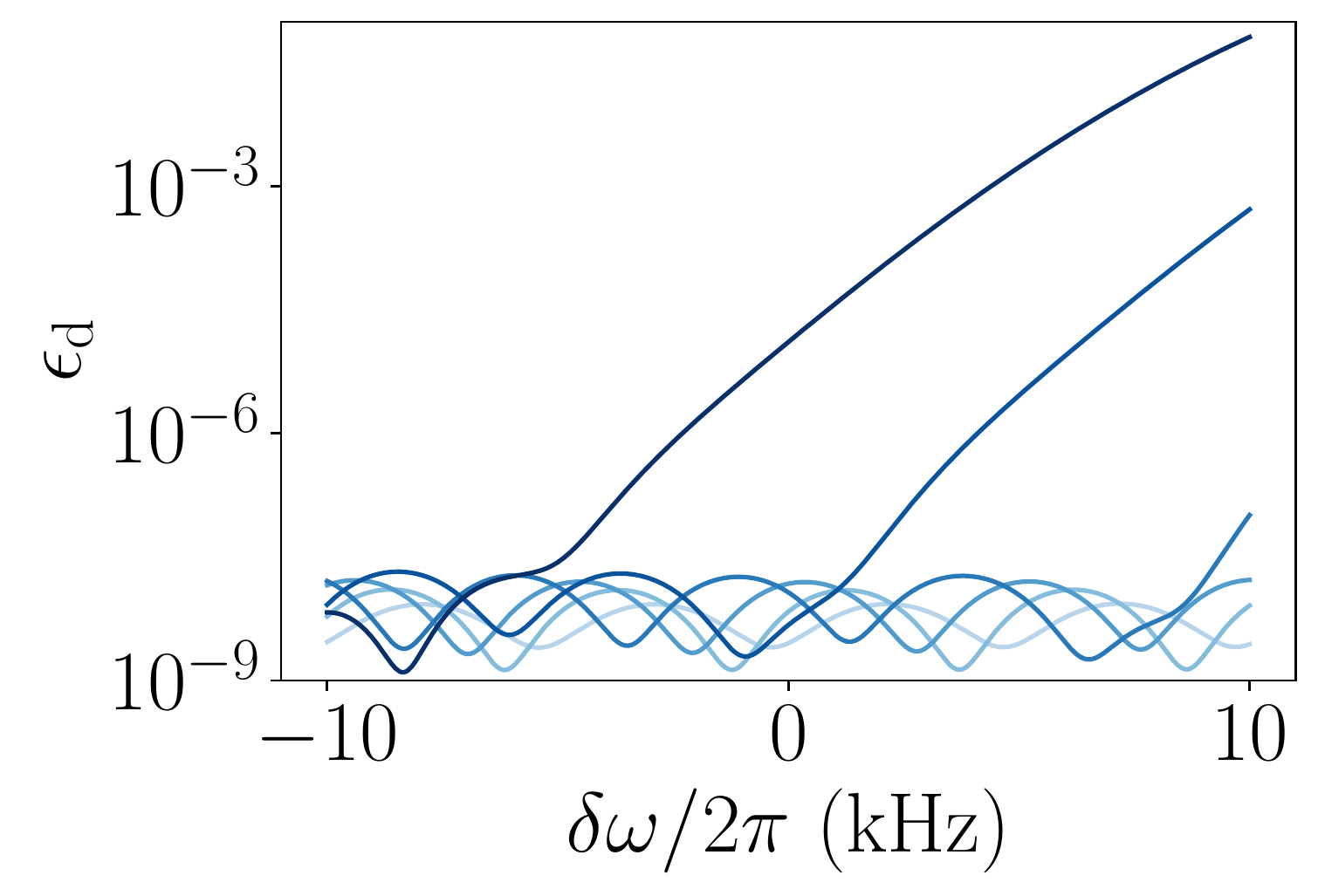}}
    \vspace{-3em}
    \caption{\raggedright}
    \end{subfigure}
    \caption{\raggedright (a and c) The rotation-angle error $\epsilon_r$ and (b and d) the displacement error $\epsilon_d$ for a center ion separation parameter of $\Delta x_0 = 3.0$\,$\mu$m. The sum of $\epsilon_r$ and $\epsilon_d$ is the state infidelity $\epsilon_s=\epsilon_d+\epsilon_r$. From light to dark, the curves correspond to (a and b) even $N$ from 2 to 32 in steps of 6 and to (c and d) odd $N$ from 3 to 33 in steps of 6. For most $N$ shown here, $\epsilon_r$ dominates as the main limiting factor for gate robustness to detuning error. For large $N$, the displacement error $\epsilon_d$ eventually lifts off the Gaussian truncation floor as the motional modes bunch closer together and significantly reduce error suppression from the factor of $e^{-\delta^2_k z^2}$.}
    \label{fig:large_N}
\end{figure}

Figure~\ref{fig:large_N} shows the contributions $\epsilon_r$ and $\epsilon_d$ to the state infidelity $\epsilon_s=\epsilon_d + \epsilon_r$ over an experimentally relevant range of frequency errors $\delta\omega/2\pi$ for ion chains of length $N=2$ to $N=33$ with $\Delta x_0=3$\,$\mu$m. Here, we have reduced the value of $\Delta x_0$ from the previous sections to achieve frequency-robust gates for all $N$. We can see that the robustness to $\delta\omega$ generally decreases with increasing $N$, but the gate error has a slightly different dependence on $\delta\omega$ for even and odd $N$. This difference comes from the Lamb-Dicke parameters of the lowest two modes for each chain, which cause the value of $\delta_c$ that solves equation~(\ref{eq:dtheta}) to lie much closer to the second lowest mode ($k=1$) for even $N$ than for odd $N$. We can see that while the robustness of $\epsilon_r$ is approximately equal for similar length chains of even and odd $N$, the smaller value of $\delta_1$ for even $N$ causes $\epsilon_d$ over this range of $\delta\omega$ to be dramatically larger than for odd $N$. 

Figure~\ref{fig:large_N} also shows that $\epsilon_r$ strongly dominates $\epsilon_d$ when $N\lesssim 20$ (for even $N$) and when $N\lesssim 27$ (for odd $N$). For $N$ greater than these values, $\delta_1$ becomes small enough that $\epsilon_d$ makes a significant contribution to $\epsilon_s$. Consequently, although $\epsilon_r$ is minimized at $\delta\omega=0$, the minimum of $\epsilon_s$ vs. $\delta\omega$ is shifted away from $\delta\omega = 0$ as $N$ is increased. This effect is consistent with our gate design since we chose to maximize the robustness of $\epsilon_r$ with respect to $\delta\omega$ instead of minimizing $\epsilon_s$. Although the sensitivity to frequency error grows with increasing $N$, the balanced Gaussian gate achieves a remarkable robustness to frequency error with $\epsilon_s$ remaining below the $10^{-2}$ level for $\delta\omega=\pm10$\,kHz for ion chains of length $N=2$ to 33.

\begin{figure}
    \raggedright
    \begin{subfigure}[b]{0.235\textwidth}
    \raggedright
    \resizebox{1.0\textwidth}{!}{
    \includegraphics{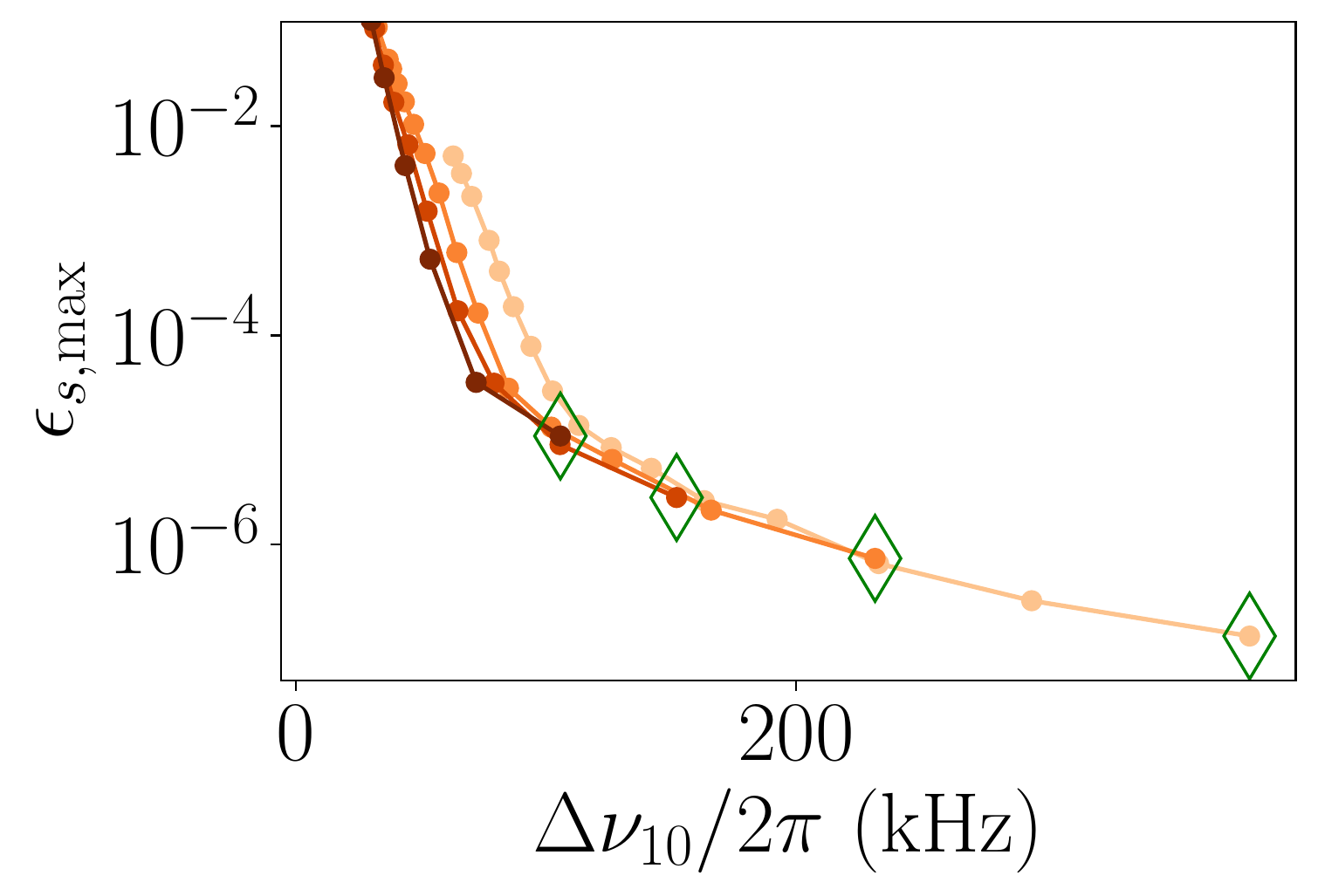}}
    \vspace{-3em}
    \caption{\raggedright}
    \end{subfigure}
        \hfill
    \begin{subfigure}[b]{0.235\textwidth}
    \raggedright
    \resizebox{1.0\textwidth}{!}{
    \includegraphics{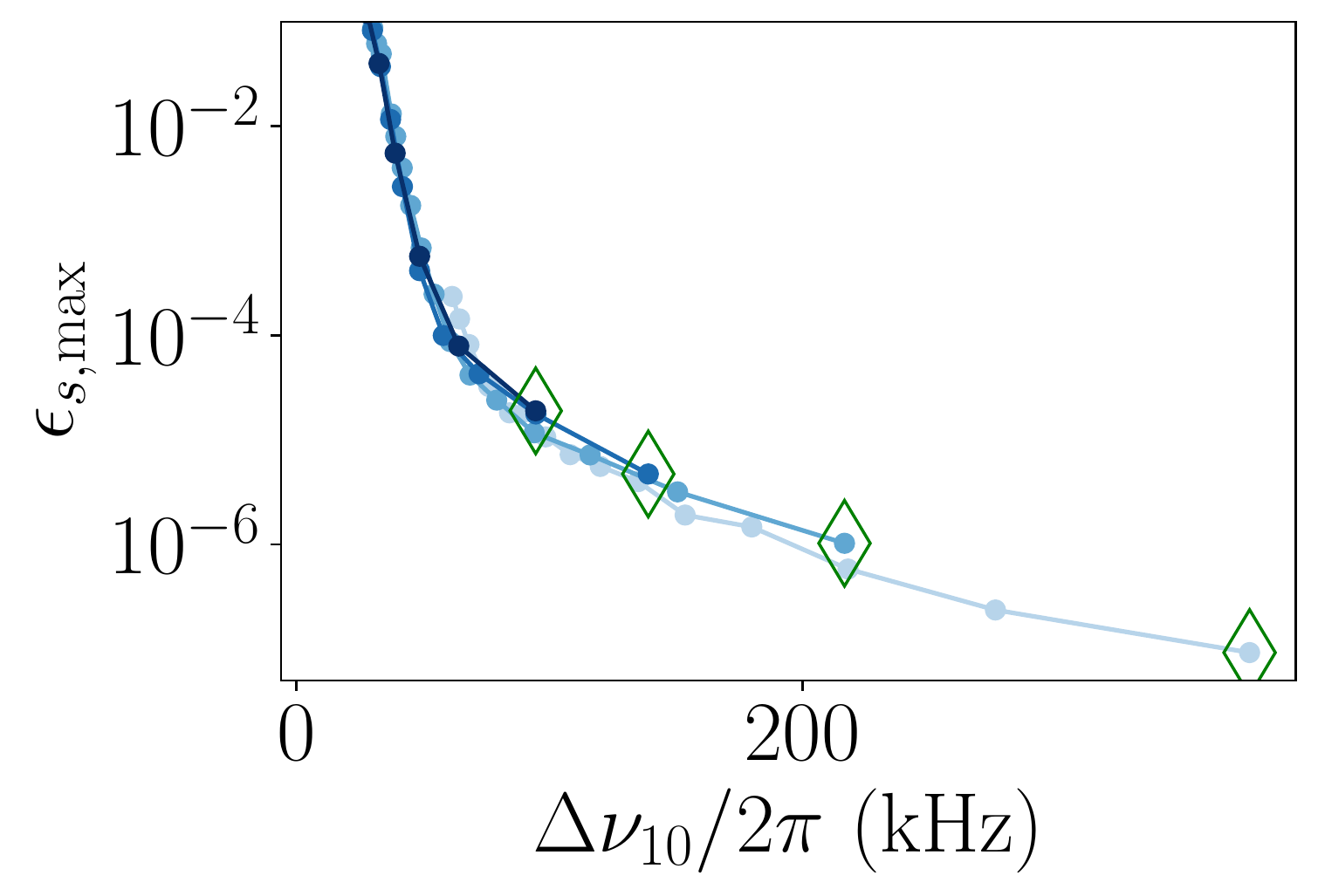}}
    \vspace{-3em}
    \caption{\raggedright}
    \end{subfigure}
    \caption{\raggedright
    The sensitivity to frequency error, as quantified by the maximum state infidelity $\epsilon_{s, \text{max}}$ over a range of $\pm 3$\,kHz from the minimum of $\epsilon_s$ vs. $\delta\omega$ for (a) even $N$ from 2 to 32 and for (b) odd $N$ from 3 to 33. The motional frequency splitting $\Delta\nu_{10}$ decreases monotonically with increasing $N$. From light to dark in each plot, the center ion separation parameter is $\Delta x_0 = 3$\,$\mu$m to $4.5$\,$\mu$m in steps of $0.5$\,$\mu$m. The green diamonds indicate (a) $N=2$ and (b) $N=3$ for each $\Delta x_0$.
    These plots show that $\Delta \nu_{10}$ is a good predictor of the sensitivity to motional frequency error for the range of parameters considered in this work.
    }
    \label{fig:max_infid}
\end{figure}

To quantify the sensitivity to frequency error for each length of chain, we compute the maximum state infidelity $\epsilon_{s, \text{max}}$ over a $\pm3$\,kHz range from the frequency that minimizes $\epsilon_s$. 
Figure~\ref{fig:max_infid} shows this measure of sensitivity for even and odd $N$, separately, and for $\Delta x_0 = 3.5\,\mu$m to $4.5\,\mu$m. For both even and odd $N$, $\epsilon_{s, \text{max}}$ is approximately a function of only the splitting between the lowest two modes $\Delta \nu_{10}$ for each value of $\Delta x_0$, despite the different motional frequencies, Lamb-Dicke parameters, number of ions $N$, and pulse shape parameters $\delta_c$ and $\Omega_0$.
Since $\Delta \nu_{10}$ appears to be a good predictor of balanced Gaussian gate performance for the chains considered in this work, we note that designs of new experiments intending to implement these gates should consider the mode spacing when choosing a value of $\Delta x_0$ and $N$.
The value of $\Delta \nu_{10}$ decreases monotonically with larger $N$ (for fixed $\Delta x_0$) and increases for smaller $\Delta x_0$ (at fixed $N$).

The approximate correspondence between $\epsilon_{s, \text{max}}$ and $\Delta \nu_{10}$ arises because $\epsilon_s$ is often dominated
by the contribution from $\epsilon_r$ for the chains we have considered.
When $\epsilon_r$ dominates, the sensitivity depends strongly on the magnitude of
$\mathrm{d}^2\theta/\mathrm{d}\delta\omega^2$ at $\delta\omega=0$, which is set by $\Delta\nu_{10}$. 
This approximate correspondence breaks down when $\epsilon_d$ becomes significant, as $\epsilon_d$ is set by the detuning $\delta_k$ instead of $\Delta \nu_{10}$. In such scenarios, $\epsilon_{s, \text{max}}$ can be substantially different for chains with different $\Delta x_0$ but the same $\Delta \nu_{10}$. For example, for $\Delta x_0 = 3\,\mu$m, the lightest curves in Fig.~\ref{fig:max_infid}, we find increased $\epsilon_{s, \text{max}}$ for long chains (low $\Delta \nu_{10}$) as compared to chains with $\Delta x_0 = 3.5\,\mu$m to $4.5\,\mu$m for the same values of $\Delta \nu_{10}$. This effect is much more pronounced for even $N$ than for odd $N$, consistent with the greater contribution from $\epsilon_d$ for even $N$. 


In this extension to larger $N$, we have reduced the ion separation parameter $\Delta x_0$ from our experimental implementation for $N=3$ to increase $\Delta\nu_{10}$ and improve the performance of MS gates at experimentally relevant frequency errors for large $N$. Although this improves the robustness of our simulated gates, a reduction in $\Delta x_0$ could have detrimental effects on quantum circuit performance that are not included in our simulations by, for example, increasing the level of crosstalk between neighboring qubits~\cite{crain:2014}. However, there has been promising research on applying spin-echo pulses to cancel out the effects of crosstalk when individual ions cannot be perfectly resolved~\cite{fang:2022}. Alternatively, one could decrease the radial trap frequency to increase $\Delta\nu_{10}$, but one needs to balance this increased robustness to frequency noise with the additional amount of anomalous heating experienced by lower frequency motional modes.
\vspace{2.5em}
\section{Conclusion and Outlook}

In summary, we have designed an MS gate that shows broad robustness to motional frequency errors, addressing a key error mechanism in trapped ion entangling gates. Further, our gate is performed with a simple analytic pulse shape, and therefore requires little in the way of computational overhead and technical complexity. In conjunction with this low cost, the prospects for scaling to a larger number of ions present a promising outlook for implementation of this gate on contemporary and next generation trapped ion systems.


\section{Acknowledgments}
This research was supported by the U.S. Department of Energy, Office of Science, Office of Advanced Scientific Computing Research Quantum Testbed Program.
Sandia National Laboratories is a multimission laboratory managed and operated by National Technology \& Engineering Solutions of Sandia, LLC, a wholly owned subsidiary of Honeywell International Inc., for the U.S. Department of Energy's National Nuclear Security Administration under contract DE-NA0003525.  This paper describes objective technical results and analysis. Any subjective views or opinions that might be expressed in the paper do not necessarily represent the views of the U.S. Department of Energy or the United States Government. 

B.P.R. and M.N.H.C. contributed equally to this work.

\bibliography{ion_refs}

\begin{thebibliography}{39}%
\makeatletter
\providecommand \@ifxundefined [1]{%
 \@ifx{#1\undefined}
}%
\providecommand \@ifnum [1]{%
 \ifnum #1\expandafter \@firstoftwo
 \else \expandafter \@secondoftwo
 \fi
}%
\providecommand \@ifx [1]{%
 \ifx #1\expandafter \@firstoftwo
 \else \expandafter \@secondoftwo
 \fi
}%
\providecommand \natexlab [1]{#1}%
\providecommand \enquote  [1]{``#1''}%
\providecommand \bibnamefont  [1]{#1}%
\providecommand \bibfnamefont [1]{#1}%
\providecommand \citenamefont [1]{#1}%
\providecommand \href@noop [0]{\@secondoftwo}%
\providecommand \href [0]{\begingroup \@sanitize@url \@href}%
\providecommand \@href[1]{\@@startlink{#1}\@@href}%
\providecommand \@@href[1]{\endgroup#1\@@endlink}%
\providecommand \@sanitize@url [0]{\catcode `\\12\catcode `\$12\catcode
  `\&12\catcode `\#12\catcode `\^12\catcode `\_12\catcode `\%12\relax}%
\providecommand \@@startlink[1]{}%
\providecommand \@@endlink[0]{}%
\providecommand \url  [0]{\begingroup\@sanitize@url \@url }%
\providecommand \@url [1]{\endgroup\@href {#1}{\urlprefix }}%
\providecommand \urlprefix  [0]{URL }%
\providecommand \Eprint [0]{\href }%
\providecommand \doibase [0]{https://doi.org/}%
\providecommand \selectlanguage [0]{\@gobble}%
\providecommand \bibinfo  [0]{\@secondoftwo}%
\providecommand \bibfield  [0]{\@secondoftwo}%
\providecommand \translation [1]{[#1]}%
\providecommand \BibitemOpen [0]{}%
\providecommand \bibitemStop [0]{}%
\providecommand \bibitemNoStop [0]{.\EOS\space}%
\providecommand \EOS [0]{\spacefactor3000\relax}%
\providecommand \BibitemShut  [1]{\csname bibitem#1\endcsname}%
\let\auto@bib@innerbib\@empty
\bibitem [{\citenamefont {S\o{}rensen}\ and\ \citenamefont
  {M\o{}lmer}(2000)}]{sorensen:2000}%
  \BibitemOpen
  \bibfield  {author} {\bibinfo {author} {\bibfnamefont {A.}~\bibnamefont
  {S\o{}rensen}}\ and\ \bibinfo {author} {\bibfnamefont {K.}~\bibnamefont
  {M\o{}lmer}},\ }\href {https://doi.org/10.1103/PhysRevA.62.022311} {\bibfield
   {journal} {\bibinfo  {journal} {Phys. Rev. A}\ }\textbf {\bibinfo {volume}
  {62}},\ \bibinfo {pages} {022311} (\bibinfo {year} {2000})}\BibitemShut
  {NoStop}%
\bibitem [{\citenamefont {Ballance}\ \emph {et~al.}(2016)\citenamefont
  {Ballance}, \citenamefont {Harty}, \citenamefont {Linke}, \citenamefont
  {Sepiol},\ and\ \citenamefont {Lucas}}]{ballance:2016}%
  \BibitemOpen
  \bibfield  {author} {\bibinfo {author} {\bibfnamefont {C.~J.}\ \bibnamefont
  {Ballance}}, \bibinfo {author} {\bibfnamefont {T.~P.}\ \bibnamefont {Harty}},
  \bibinfo {author} {\bibfnamefont {N.~M.}\ \bibnamefont {Linke}}, \bibinfo
  {author} {\bibfnamefont {M.~A.}\ \bibnamefont {Sepiol}},\ and\ \bibinfo
  {author} {\bibfnamefont {D.~M.}\ \bibnamefont {Lucas}},\ }\href
  {https://doi.org/10.1103/PhysRevLett.117.060504} {\bibfield  {journal}
  {\bibinfo  {journal} {Phys. Rev. Lett.}\ }\textbf {\bibinfo {volume} {117}},\
  \bibinfo {pages} {060504} (\bibinfo {year} {2016})}\BibitemShut {NoStop}%
\bibitem [{\citenamefont {Gaebler}\ \emph {et~al.}(2016)\citenamefont
  {Gaebler}, \citenamefont {Tan}, \citenamefont {Lin}, \citenamefont {Wan},
  \citenamefont {Bowler}, \citenamefont {Keith}, \citenamefont {Glancy},
  \citenamefont {Coakley}, \citenamefont {Knill}, \citenamefont {Leibfried},\
  and\ \citenamefont {Wineland}}]{gaebler:2016}%
  \BibitemOpen
  \bibfield  {author} {\bibinfo {author} {\bibfnamefont {J.~P.}\ \bibnamefont
  {Gaebler}}, \bibinfo {author} {\bibfnamefont {T.~R.}\ \bibnamefont {Tan}},
  \bibinfo {author} {\bibfnamefont {Y.}~\bibnamefont {Lin}}, \bibinfo {author}
  {\bibfnamefont {Y.}~\bibnamefont {Wan}}, \bibinfo {author} {\bibfnamefont
  {R.}~\bibnamefont {Bowler}}, \bibinfo {author} {\bibfnamefont {A.~C.}\
  \bibnamefont {Keith}}, \bibinfo {author} {\bibfnamefont {S.}~\bibnamefont
  {Glancy}}, \bibinfo {author} {\bibfnamefont {K.}~\bibnamefont {Coakley}},
  \bibinfo {author} {\bibfnamefont {E.}~\bibnamefont {Knill}}, \bibinfo
  {author} {\bibfnamefont {D.}~\bibnamefont {Leibfried}},\ and\ \bibinfo
  {author} {\bibfnamefont {D.~J.}\ \bibnamefont {Wineland}},\ }\href
  {https://doi.org/10.1103/PhysRevLett.117.060505} {\bibfield  {journal}
  {\bibinfo  {journal} {Phys. Rev. Lett.}\ }\textbf {\bibinfo {volume} {117}},\
  \bibinfo {pages} {060505} (\bibinfo {year} {2016})}\BibitemShut {NoStop}%
\bibitem [{\citenamefont {Lanyon}\ \emph {et~al.}(2011)\citenamefont {Lanyon},
  \citenamefont {Hempel}, \citenamefont {Nigg}, \citenamefont {Müller},
  \citenamefont {Gerritsma}, \citenamefont {Zähringer}, \citenamefont
  {Schindler}, \citenamefont {Barreiro}, \citenamefont {Rambach}, \citenamefont
  {Kirchmair}, \citenamefont {Hennrich}, \citenamefont {Zoller}, \citenamefont
  {Blatt},\ and\ \citenamefont {Roos}}]{lanyon:2011}%
  \BibitemOpen
  \bibfield  {author} {\bibinfo {author} {\bibfnamefont {B.~P.}\ \bibnamefont
  {Lanyon}}, \bibinfo {author} {\bibfnamefont {C.}~\bibnamefont {Hempel}},
  \bibinfo {author} {\bibfnamefont {D.}~\bibnamefont {Nigg}}, \bibinfo {author}
  {\bibfnamefont {M.}~\bibnamefont {Müller}}, \bibinfo {author} {\bibfnamefont
  {R.}~\bibnamefont {Gerritsma}}, \bibinfo {author} {\bibfnamefont
  {F.}~\bibnamefont {Zähringer}}, \bibinfo {author} {\bibfnamefont
  {P.}~\bibnamefont {Schindler}}, \bibinfo {author} {\bibfnamefont {J.~T.}\
  \bibnamefont {Barreiro}}, \bibinfo {author} {\bibfnamefont {M.}~\bibnamefont
  {Rambach}}, \bibinfo {author} {\bibfnamefont {G.}~\bibnamefont {Kirchmair}},
  \bibinfo {author} {\bibfnamefont {M.}~\bibnamefont {Hennrich}}, \bibinfo
  {author} {\bibfnamefont {P.}~\bibnamefont {Zoller}}, \bibinfo {author}
  {\bibfnamefont {R.}~\bibnamefont {Blatt}},\ and\ \bibinfo {author}
  {\bibfnamefont {C.~F.}\ \bibnamefont {Roos}},\ }\href
  {https://doi.org/10.1126/science.1208001} {\bibfield  {journal} {\bibinfo
  {journal} {Science}\ }\textbf {\bibinfo {volume} {334}},\ \bibinfo {pages}
  {57} (\bibinfo {year} {2011})}\BibitemShut {NoStop}%
\bibitem [{\citenamefont {Shor}(1995)}]{shor:1995}%
  \BibitemOpen
  \bibfield  {author} {\bibinfo {author} {\bibfnamefont {P.~W.}\ \bibnamefont
  {Shor}},\ }\href {https://doi.org/10.1103/PhysRevA.52.R2493} {\bibfield
  {journal} {\bibinfo  {journal} {Phys. Rev. A}\ }\textbf {\bibinfo {volume}
  {52}},\ \bibinfo {pages} {R2493} (\bibinfo {year} {1995})}\BibitemShut
  {NoStop}%
\bibitem [{\citenamefont {Kitaev}(2003)}]{kitaev:2003}%
  \BibitemOpen
  \bibfield  {author} {\bibinfo {author} {\bibfnamefont {A.}~\bibnamefont
  {Kitaev}},\ }\href
  {https://doi.org/https://doi.org/10.1016/S0003-4916(02)00018-0} {\bibfield
  {journal} {\bibinfo  {journal} {Annals of Physics}\ }\textbf {\bibinfo
  {volume} {303}},\ \bibinfo {pages} {2} (\bibinfo {year} {2003})}\BibitemShut
  {NoStop}%
\bibitem [{\citenamefont {Aharonov}\ and\ \citenamefont
  {Ben-Or}(2008)}]{aharonov:2008}%
  \BibitemOpen
  \bibfield  {author} {\bibinfo {author} {\bibfnamefont {D.}~\bibnamefont
  {Aharonov}}\ and\ \bibinfo {author} {\bibfnamefont {M.}~\bibnamefont
  {Ben-Or}},\ }\href {https://doi.org/10.1137/S0097539799359385} {\bibfield
  {journal} {\bibinfo  {journal} {SIAM Journal on Computing}\ }\textbf
  {\bibinfo {volume} {38}},\ \bibinfo {pages} {1207} (\bibinfo {year}
  {2008})}\BibitemShut {NoStop}%
\bibitem [{\citenamefont {Debnath}\ \emph {et~al.}(2016)\citenamefont
  {Debnath}, \citenamefont {Linke}, \citenamefont {Figgatt}, \citenamefont
  {Landsman}, \citenamefont {Wright},\ and\ \citenamefont
  {Monroe}}]{debnath:2016}%
  \BibitemOpen
  \bibfield  {author} {\bibinfo {author} {\bibfnamefont {S.}~\bibnamefont
  {Debnath}}, \bibinfo {author} {\bibfnamefont {N.~M.}\ \bibnamefont {Linke}},
  \bibinfo {author} {\bibfnamefont {C.}~\bibnamefont {Figgatt}}, \bibinfo
  {author} {\bibfnamefont {K.~A.}\ \bibnamefont {Landsman}}, \bibinfo {author}
  {\bibfnamefont {K.}~\bibnamefont {Wright}},\ and\ \bibinfo {author}
  {\bibfnamefont {C.}~\bibnamefont {Monroe}},\ }\href
  {https://doi.org/10.1038/nature18648} {\bibfield  {journal} {\bibinfo
  {journal} {Nature}\ }\textbf {\bibinfo {volume} {536}},\ \bibinfo {pages}
  {63} (\bibinfo {year} {2016})}\BibitemShut {NoStop}%
\bibitem [{\citenamefont {Wright}\ \emph {et~al.}(2019)\citenamefont {Wright},
  \citenamefont {Beck}, \citenamefont {Debnath}, \citenamefont {Amini},
  \citenamefont {Nam}, \citenamefont {Grzesiak}, \citenamefont {Chen},
  \citenamefont {Pisenti}, \citenamefont {Chmielewski}, \citenamefont
  {Collins}, \citenamefont {Hudek}, \citenamefont {Mizrahi}, \citenamefont
  {Wong-Campos}, \citenamefont {Allen}, \citenamefont {Apisdorf}, \citenamefont
  {Solomon}, \citenamefont {Williams}, \citenamefont {Ducore}, \citenamefont
  {Blinov}, \citenamefont {Kreikemeier}, \citenamefont {Chaplin}, \citenamefont
  {Keesan}, \citenamefont {Monroe},\ and\ \citenamefont {Kim}}]{wright:2019}%
  \BibitemOpen
  \bibfield  {author} {\bibinfo {author} {\bibfnamefont {K.}~\bibnamefont
  {Wright}}, \bibinfo {author} {\bibfnamefont {K.~M.}\ \bibnamefont {Beck}},
  \bibinfo {author} {\bibfnamefont {S.}~\bibnamefont {Debnath}}, \bibinfo
  {author} {\bibfnamefont {J.~M.}\ \bibnamefont {Amini}}, \bibinfo {author}
  {\bibfnamefont {Y.}~\bibnamefont {Nam}}, \bibinfo {author} {\bibfnamefont
  {N.}~\bibnamefont {Grzesiak}}, \bibinfo {author} {\bibfnamefont {J.-S.}\
  \bibnamefont {Chen}}, \bibinfo {author} {\bibfnamefont {N.~C.}\ \bibnamefont
  {Pisenti}}, \bibinfo {author} {\bibfnamefont {M.}~\bibnamefont
  {Chmielewski}}, \bibinfo {author} {\bibfnamefont {C.}~\bibnamefont
  {Collins}}, \bibinfo {author} {\bibfnamefont {K.~M.}\ \bibnamefont {Hudek}},
  \bibinfo {author} {\bibfnamefont {J.}~\bibnamefont {Mizrahi}}, \bibinfo
  {author} {\bibfnamefont {J.~D.}\ \bibnamefont {Wong-Campos}}, \bibinfo
  {author} {\bibfnamefont {S.}~\bibnamefont {Allen}}, \bibinfo {author}
  {\bibfnamefont {J.}~\bibnamefont {Apisdorf}}, \bibinfo {author}
  {\bibfnamefont {P.}~\bibnamefont {Solomon}}, \bibinfo {author} {\bibfnamefont
  {M.}~\bibnamefont {Williams}}, \bibinfo {author} {\bibfnamefont {A.~M.}\
  \bibnamefont {Ducore}}, \bibinfo {author} {\bibfnamefont {A.}~\bibnamefont
  {Blinov}}, \bibinfo {author} {\bibfnamefont {S.~M.}\ \bibnamefont
  {Kreikemeier}}, \bibinfo {author} {\bibfnamefont {V.}~\bibnamefont
  {Chaplin}}, \bibinfo {author} {\bibfnamefont {M.}~\bibnamefont {Keesan}},
  \bibinfo {author} {\bibfnamefont {C.}~\bibnamefont {Monroe}},\ and\ \bibinfo
  {author} {\bibfnamefont {J.}~\bibnamefont {Kim}},\ }\href
  {https://doi.org/10.1038/s41467-019-13534-2} {\bibfield  {journal} {\bibinfo
  {journal} {Nature Communications}\ }\textbf {\bibinfo {volume} {10}},\
  \bibinfo {pages} {5464} (\bibinfo {year} {2019})}\BibitemShut {NoStop}%
\bibitem [{\citenamefont {Zhu}\ \emph {et~al.}(2006)\citenamefont {Zhu},
  \citenamefont {Monroe},\ and\ \citenamefont {Duan}}]{zhu:2006}%
  \BibitemOpen
  \bibfield  {author} {\bibinfo {author} {\bibfnamefont {S.-L.}\ \bibnamefont
  {Zhu}}, \bibinfo {author} {\bibfnamefont {C.}~\bibnamefont {Monroe}},\ and\
  \bibinfo {author} {\bibfnamefont {L.-M.}\ \bibnamefont {Duan}},\ }\href
  {https://doi.org/10.1209/epl/i2005-10424-4} {\bibfield  {journal} {\bibinfo
  {journal} {Europhysics Letters ({EPL})}\ }\textbf {\bibinfo {volume} {73}},\
  \bibinfo {pages} {485} (\bibinfo {year} {2006})}\BibitemShut {NoStop}%
\bibitem [{\citenamefont {Roos}(2008)}]{Roos2008}%
  \BibitemOpen
  \bibfield  {author} {\bibinfo {author} {\bibfnamefont {C.~F.}\ \bibnamefont
  {Roos}},\ }\href {https://doi.org/10.1088/1367-2630/10/1/013002} {\bibfield
  {journal} {\bibinfo  {journal} {New Journal of Physics}\ }\textbf {\bibinfo
  {volume} {10}},\ \bibinfo {pages} {013002} (\bibinfo {year}
  {2008})}\BibitemShut {NoStop}%
\bibitem [{\citenamefont {Benhelm}\ \emph {et~al.}(2008)\citenamefont
  {Benhelm}, \citenamefont {Kirchmair}, \citenamefont {Roos},\ and\
  \citenamefont {Blatt}}]{BenhelmRoos2008}%
  \BibitemOpen
  \bibfield  {author} {\bibinfo {author} {\bibfnamefont {J.}~\bibnamefont
  {Benhelm}}, \bibinfo {author} {\bibfnamefont {G.}~\bibnamefont {Kirchmair}},
  \bibinfo {author} {\bibfnamefont {C.~F.}\ \bibnamefont {Roos}},\ and\
  \bibinfo {author} {\bibfnamefont {R.}~\bibnamefont {Blatt}},\ }\href
  {https://doi.org/10.1038/nphys961} {\bibfield  {journal} {\bibinfo  {journal}
  {Nature Physics}\ }\textbf {\bibinfo {volume} {4}},\ \bibinfo {pages} {463}
  (\bibinfo {year} {2008})}\BibitemShut {NoStop}%
\bibitem [{\citenamefont {Choi}\ \emph {et~al.}(2014)\citenamefont {Choi},
  \citenamefont {Debnath}, \citenamefont {Manning}, \citenamefont {Figgatt},
  \citenamefont {Gong}, \citenamefont {Duan},\ and\ \citenamefont
  {Monroe}}]{choi:2014}%
  \BibitemOpen
  \bibfield  {author} {\bibinfo {author} {\bibfnamefont {T.}~\bibnamefont
  {Choi}}, \bibinfo {author} {\bibfnamefont {S.}~\bibnamefont {Debnath}},
  \bibinfo {author} {\bibfnamefont {T.~A.}\ \bibnamefont {Manning}}, \bibinfo
  {author} {\bibfnamefont {C.}~\bibnamefont {Figgatt}}, \bibinfo {author}
  {\bibfnamefont {Z.-X.}\ \bibnamefont {Gong}}, \bibinfo {author}
  {\bibfnamefont {L.-M.}\ \bibnamefont {Duan}},\ and\ \bibinfo {author}
  {\bibfnamefont {C.}~\bibnamefont {Monroe}},\ }\href
  {https://doi.org/10.1103/PhysRevLett.112.190502} {\bibfield  {journal}
  {\bibinfo  {journal} {Phys. Rev. Lett.}\ }\textbf {\bibinfo {volume} {112}},\
  \bibinfo {pages} {190502} (\bibinfo {year} {2014})}\BibitemShut {NoStop}%
\bibitem [{\citenamefont {Tinkey}\ \emph {et~al.}(2022)\citenamefont {Tinkey},
  \citenamefont {Clark}, \citenamefont {Sawyer},\ and\ \citenamefont
  {Brown}}]{tinkey:2022}%
  \BibitemOpen
  \bibfield  {author} {\bibinfo {author} {\bibfnamefont {H.~N.}\ \bibnamefont
  {Tinkey}}, \bibinfo {author} {\bibfnamefont {C.~R.}\ \bibnamefont {Clark}},
  \bibinfo {author} {\bibfnamefont {B.~C.}\ \bibnamefont {Sawyer}},\ and\
  \bibinfo {author} {\bibfnamefont {K.~R.}\ \bibnamefont {Brown}},\ }\href
  {https://doi.org/10.1103/PhysRevLett.128.050502} {\bibfield  {journal}
  {\bibinfo  {journal} {Phys. Rev. Lett.}\ }\textbf {\bibinfo {volume} {128}},\
  \bibinfo {pages} {050502} (\bibinfo {year} {2022})}\BibitemShut {NoStop}%
\bibitem [{\citenamefont {Leung}\ \emph {et~al.}(2018)\citenamefont {Leung},
  \citenamefont {Landsman}, \citenamefont {Figgatt}, \citenamefont {Linke},
  \citenamefont {Monroe},\ and\ \citenamefont {Brown}}]{Leung5ion}%
  \BibitemOpen
  \bibfield  {author} {\bibinfo {author} {\bibfnamefont {P.~H.}\ \bibnamefont
  {Leung}}, \bibinfo {author} {\bibfnamefont {K.~A.}\ \bibnamefont {Landsman}},
  \bibinfo {author} {\bibfnamefont {C.}~\bibnamefont {Figgatt}}, \bibinfo
  {author} {\bibfnamefont {N.~M.}\ \bibnamefont {Linke}}, \bibinfo {author}
  {\bibfnamefont {C.}~\bibnamefont {Monroe}},\ and\ \bibinfo {author}
  {\bibfnamefont {K.~R.}\ \bibnamefont {Brown}},\ }\href
  {https://doi.org/10.1103/PhysRevLett.120.020501} {\bibfield  {journal}
  {\bibinfo  {journal} {Phys. Rev. Lett.}\ }\textbf {\bibinfo {volume} {120}},\
  \bibinfo {pages} {020501} (\bibinfo {year} {2018})}\BibitemShut {NoStop}%
\bibitem [{\citenamefont {Wang}\ \emph {et~al.}(2020)\citenamefont {Wang},
  \citenamefont {Crain}, \citenamefont {Fang}, \citenamefont {Zhang},
  \citenamefont {Huang}, \citenamefont {Liang}, \citenamefont {Leung},
  \citenamefont {Brown},\ and\ \citenamefont {Kim}}]{wang:2020}%
  \BibitemOpen
  \bibfield  {author} {\bibinfo {author} {\bibfnamefont {Y.}~\bibnamefont
  {Wang}}, \bibinfo {author} {\bibfnamefont {S.}~\bibnamefont {Crain}},
  \bibinfo {author} {\bibfnamefont {C.}~\bibnamefont {Fang}}, \bibinfo {author}
  {\bibfnamefont {B.}~\bibnamefont {Zhang}}, \bibinfo {author} {\bibfnamefont
  {S.}~\bibnamefont {Huang}}, \bibinfo {author} {\bibfnamefont
  {Q.}~\bibnamefont {Liang}}, \bibinfo {author} {\bibfnamefont {P.~H.}\
  \bibnamefont {Leung}}, \bibinfo {author} {\bibfnamefont {K.~R.}\ \bibnamefont
  {Brown}},\ and\ \bibinfo {author} {\bibfnamefont {J.}~\bibnamefont {Kim}},\
  }\href {https://doi.org/10.1103/PhysRevLett.125.150505} {\bibfield  {journal}
  {\bibinfo  {journal} {Phys. Rev. Lett.}\ }\textbf {\bibinfo {volume} {125}},\
  \bibinfo {pages} {150505} (\bibinfo {year} {2020})}\BibitemShut {NoStop}%
\bibitem [{\citenamefont {Leung}\ and\ \citenamefont
  {Brown}(2018)}]{leung:2018}%
  \BibitemOpen
  \bibfield  {author} {\bibinfo {author} {\bibfnamefont {P.~H.}\ \bibnamefont
  {Leung}}\ and\ \bibinfo {author} {\bibfnamefont {K.~R.}\ \bibnamefont
  {Brown}},\ }\href {https://doi.org/10.1103/PhysRevA.98.032318} {\bibfield
  {journal} {\bibinfo  {journal} {Phys. Rev. A}\ }\textbf {\bibinfo {volume}
  {98}},\ \bibinfo {pages} {032318} (\bibinfo {year} {2018})}\BibitemShut
  {NoStop}%
\bibitem [{\citenamefont {Landsman}\ \emph {et~al.}(2019)\citenamefont
  {Landsman}, \citenamefont {Wu}, \citenamefont {Leung}, \citenamefont {Zhu},
  \citenamefont {Linke}, \citenamefont {Brown}, \citenamefont {Duan},\ and\
  \citenamefont {Monroe}}]{landsman:2019}%
  \BibitemOpen
  \bibfield  {author} {\bibinfo {author} {\bibfnamefont {K.~A.}\ \bibnamefont
  {Landsman}}, \bibinfo {author} {\bibfnamefont {Y.}~\bibnamefont {Wu}},
  \bibinfo {author} {\bibfnamefont {P.~H.}\ \bibnamefont {Leung}}, \bibinfo
  {author} {\bibfnamefont {D.}~\bibnamefont {Zhu}}, \bibinfo {author}
  {\bibfnamefont {N.~M.}\ \bibnamefont {Linke}}, \bibinfo {author}
  {\bibfnamefont {K.~R.}\ \bibnamefont {Brown}}, \bibinfo {author}
  {\bibfnamefont {L.}~\bibnamefont {Duan}},\ and\ \bibinfo {author}
  {\bibfnamefont {C.}~\bibnamefont {Monroe}},\ }\href
  {https://doi.org/10.1103/PhysRevA.100.022332} {\bibfield  {journal} {\bibinfo
   {journal} {Phys. Rev. A}\ }\textbf {\bibinfo {volume} {100}},\ \bibinfo
  {pages} {022332} (\bibinfo {year} {2019})}\BibitemShut {NoStop}%
\bibitem [{\citenamefont {Green}\ and\ \citenamefont
  {Biercuk}(2015)}]{green:2015}%
  \BibitemOpen
  \bibfield  {author} {\bibinfo {author} {\bibfnamefont {T.~J.}\ \bibnamefont
  {Green}}\ and\ \bibinfo {author} {\bibfnamefont {M.~J.}\ \bibnamefont
  {Biercuk}},\ }\href {https://doi.org/10.1103/PhysRevLett.114.120502}
  {\bibfield  {journal} {\bibinfo  {journal} {Phys. Rev. Lett.}\ }\textbf
  {\bibinfo {volume} {114}},\ \bibinfo {pages} {120502} (\bibinfo {year}
  {2015})}\BibitemShut {NoStop}%
\bibitem [{\citenamefont {Lu}\ \emph {et~al.}(2019)\citenamefont {Lu},
  \citenamefont {Zhang}, \citenamefont {Zhang}, \citenamefont {Chen},
  \citenamefont {Shen}, \citenamefont {Zhang}, \citenamefont {Zhang},\ and\
  \citenamefont {Kim}}]{lu:2019}%
  \BibitemOpen
  \bibfield  {author} {\bibinfo {author} {\bibfnamefont {Y.}~\bibnamefont
  {Lu}}, \bibinfo {author} {\bibfnamefont {S.}~\bibnamefont {Zhang}}, \bibinfo
  {author} {\bibfnamefont {K.}~\bibnamefont {Zhang}}, \bibinfo {author}
  {\bibfnamefont {W.}~\bibnamefont {Chen}}, \bibinfo {author} {\bibfnamefont
  {Y.}~\bibnamefont {Shen}}, \bibinfo {author} {\bibfnamefont {J.}~\bibnamefont
  {Zhang}}, \bibinfo {author} {\bibfnamefont {J.-N.}\ \bibnamefont {Zhang}},\
  and\ \bibinfo {author} {\bibfnamefont {K.}~\bibnamefont {Kim}},\ }\href
  {https://doi.org/10.1038/s41586-019-1428-4} {\bibfield  {journal} {\bibinfo
  {journal} {Nature}\ }\textbf {\bibinfo {volume} {572}},\ \bibinfo {pages}
  {363} (\bibinfo {year} {2019})}\BibitemShut {NoStop}%
\bibitem [{\citenamefont {Milne}\ \emph {et~al.}(2020)\citenamefont {Milne},
  \citenamefont {Edmunds}, \citenamefont {Hempel}, \citenamefont {Roy},
  \citenamefont {Mavadia},\ and\ \citenamefont {Biercuk}}]{milne:2020}%
  \BibitemOpen
  \bibfield  {author} {\bibinfo {author} {\bibfnamefont {A.~R.}\ \bibnamefont
  {Milne}}, \bibinfo {author} {\bibfnamefont {C.~L.}\ \bibnamefont {Edmunds}},
  \bibinfo {author} {\bibfnamefont {C.}~\bibnamefont {Hempel}}, \bibinfo
  {author} {\bibfnamefont {F.}~\bibnamefont {Roy}}, \bibinfo {author}
  {\bibfnamefont {S.}~\bibnamefont {Mavadia}},\ and\ \bibinfo {author}
  {\bibfnamefont {M.~J.}\ \bibnamefont {Biercuk}},\ }\href
  {https://doi.org/10.1103/PhysRevApplied.13.024022} {\bibfield  {journal}
  {\bibinfo  {journal} {Phys. Rev. Applied}\ }\textbf {\bibinfo {volume}
  {13}},\ \bibinfo {pages} {024022} (\bibinfo {year} {2020})}\BibitemShut
  {NoStop}%
\bibitem [{\citenamefont {Kang}\ \emph {et~al.}(2021)\citenamefont {Kang},
  \citenamefont {Liang}, \citenamefont {Zhang}, \citenamefont {Huang},
  \citenamefont {Wang}, \citenamefont {Fang}, \citenamefont {Kim},\ and\
  \citenamefont {Brown}}]{KangBRobust}%
  \BibitemOpen
  \bibfield  {author} {\bibinfo {author} {\bibfnamefont {M.}~\bibnamefont
  {Kang}}, \bibinfo {author} {\bibfnamefont {Q.}~\bibnamefont {Liang}},
  \bibinfo {author} {\bibfnamefont {B.}~\bibnamefont {Zhang}}, \bibinfo
  {author} {\bibfnamefont {S.}~\bibnamefont {Huang}}, \bibinfo {author}
  {\bibfnamefont {Y.}~\bibnamefont {Wang}}, \bibinfo {author} {\bibfnamefont
  {C.}~\bibnamefont {Fang}}, \bibinfo {author} {\bibfnamefont {J.}~\bibnamefont
  {Kim}},\ and\ \bibinfo {author} {\bibfnamefont {K.~R.}\ \bibnamefont
  {Brown}},\ }\href {https://doi.org/10.1103/PhysRevApplied.16.024039}
  {\bibfield  {journal} {\bibinfo  {journal} {Phys. Rev. Applied}\ }\textbf
  {\bibinfo {volume} {16}},\ \bibinfo {pages} {024039} (\bibinfo {year}
  {2021})}\BibitemShut {NoStop}%
\bibitem [{\citenamefont {Iverson}\ and\ \citenamefont
  {Preskill}(2020)}]{iverson:2020}%
  \BibitemOpen
  \bibfield  {author} {\bibinfo {author} {\bibfnamefont {J.~K.}\ \bibnamefont
  {Iverson}}\ and\ \bibinfo {author} {\bibfnamefont {J.}~\bibnamefont
  {Preskill}},\ }\href {https://doi.org/10.1088/1367-2630/ab8e5c} {\bibfield
  {journal} {\bibinfo  {journal} {New Journal of Physics}\ }\textbf {\bibinfo
  {volume} {22}},\ \bibinfo {pages} {073066} (\bibinfo {year}
  {2020})}\BibitemShut {NoStop}%
\bibitem [{\citenamefont {Clark}\ \emph {et~al.}(2021)\citenamefont {Clark},
  \citenamefont {Lobser}, \citenamefont {Revelle}, \citenamefont {Yale},
  \citenamefont {Bossert}, \citenamefont {Burch}, \citenamefont {Chow},
  \citenamefont {Hogle}, \citenamefont {Ivory}, \citenamefont {Pehr},
  \citenamefont {Salzbrenner}, \citenamefont {Stick}, \citenamefont {Sweatt},
  \citenamefont {Wilson}, \citenamefont {Winrow},\ and\ \citenamefont
  {Maunz}}]{QSCOUTManual}%
  \BibitemOpen
  \bibfield  {author} {\bibinfo {author} {\bibfnamefont {S.~M.}\ \bibnamefont
  {Clark}}, \bibinfo {author} {\bibfnamefont {D.}~\bibnamefont {Lobser}},
  \bibinfo {author} {\bibfnamefont {M.~C.}\ \bibnamefont {Revelle}}, \bibinfo
  {author} {\bibfnamefont {C.~G.}\ \bibnamefont {Yale}}, \bibinfo {author}
  {\bibfnamefont {D.}~\bibnamefont {Bossert}}, \bibinfo {author} {\bibfnamefont
  {A.~D.}\ \bibnamefont {Burch}}, \bibinfo {author} {\bibfnamefont {M.~N.}\
  \bibnamefont {Chow}}, \bibinfo {author} {\bibfnamefont {C.~W.}\ \bibnamefont
  {Hogle}}, \bibinfo {author} {\bibfnamefont {M.}~\bibnamefont {Ivory}},
  \bibinfo {author} {\bibfnamefont {J.}~\bibnamefont {Pehr}}, \bibinfo {author}
  {\bibfnamefont {B.}~\bibnamefont {Salzbrenner}}, \bibinfo {author}
  {\bibfnamefont {D.}~\bibnamefont {Stick}}, \bibinfo {author} {\bibfnamefont
  {W.}~\bibnamefont {Sweatt}}, \bibinfo {author} {\bibfnamefont {J.~M.}\
  \bibnamefont {Wilson}}, \bibinfo {author} {\bibfnamefont {E.}~\bibnamefont
  {Winrow}},\ and\ \bibinfo {author} {\bibfnamefont {P.}~\bibnamefont
  {Maunz}},\ }\href {https://doi.org/10.1109/TQE.2021.3096480} {\bibfield
  {journal} {\bibinfo  {journal} {IEEE Transactions on Quantum Engineering}\
  }\textbf {\bibinfo {volume} {2}},\ \bibinfo {pages} {1} (\bibinfo {year}
  {2021})}\BibitemShut {NoStop}%
\bibitem [{\citenamefont {Ruzic}\ \emph {et~al.}(2022)\citenamefont {Ruzic},
  \citenamefont {Barrick}, \citenamefont {Hunker}, \citenamefont {Law},
  \citenamefont {McFarland}, \citenamefont {McGuinness}, \citenamefont
  {Parazzoli}, \citenamefont {Sterk}, \citenamefont {Van Der~Wall},\ and\
  \citenamefont {Stick}}]{ruzic:2022}%
  \BibitemOpen
  \bibfield  {author} {\bibinfo {author} {\bibfnamefont {B.~P.}\ \bibnamefont
  {Ruzic}}, \bibinfo {author} {\bibfnamefont {T.~A.}\ \bibnamefont {Barrick}},
  \bibinfo {author} {\bibfnamefont {J.~D.}\ \bibnamefont {Hunker}}, \bibinfo
  {author} {\bibfnamefont {R.~J.}\ \bibnamefont {Law}}, \bibinfo {author}
  {\bibfnamefont {B.~K.}\ \bibnamefont {McFarland}}, \bibinfo {author}
  {\bibfnamefont {H.~J.}\ \bibnamefont {McGuinness}}, \bibinfo {author}
  {\bibfnamefont {L.~P.}\ \bibnamefont {Parazzoli}}, \bibinfo {author}
  {\bibfnamefont {J.~D.}\ \bibnamefont {Sterk}}, \bibinfo {author}
  {\bibfnamefont {J.~W.}\ \bibnamefont {Van Der~Wall}},\ and\ \bibinfo {author}
  {\bibfnamefont {D.}~\bibnamefont {Stick}},\ }\href
  {https://doi.org/10.1103/PhysRevA.105.052409} {\bibfield  {journal} {\bibinfo
   {journal} {Phys. Rev. A}\ }\textbf {\bibinfo {volume} {105}},\ \bibinfo
  {pages} {052409} (\bibinfo {year} {2022})}\BibitemShut {NoStop}%
\bibitem [{\citenamefont {Bruzewicz}\ \emph {et~al.}(2015)\citenamefont
  {Bruzewicz}, \citenamefont {Sage},\ and\ \citenamefont
  {Chiaverini}}]{bruzewicz:2015}%
  \BibitemOpen
  \bibfield  {author} {\bibinfo {author} {\bibfnamefont {C.~D.}\ \bibnamefont
  {Bruzewicz}}, \bibinfo {author} {\bibfnamefont {J.~M.}\ \bibnamefont
  {Sage}},\ and\ \bibinfo {author} {\bibfnamefont {J.}~\bibnamefont
  {Chiaverini}},\ }\href {https://doi.org/10.1103/PhysRevA.91.041402}
  {\bibfield  {journal} {\bibinfo  {journal} {Phys. Rev. A}\ }\textbf {\bibinfo
  {volume} {91}},\ \bibinfo {pages} {041402} (\bibinfo {year}
  {2015})}\BibitemShut {NoStop}%
\bibitem [{\citenamefont {Boldin}\ \emph {et~al.}(2018)\citenamefont {Boldin},
  \citenamefont {Kraft},\ and\ \citenamefont {Wunderlich}}]{boldin:2018}%
  \BibitemOpen
  \bibfield  {author} {\bibinfo {author} {\bibfnamefont {I.~A.}\ \bibnamefont
  {Boldin}}, \bibinfo {author} {\bibfnamefont {A.}~\bibnamefont {Kraft}},\ and\
  \bibinfo {author} {\bibfnamefont {C.}~\bibnamefont {Wunderlich}},\ }\href
  {https://doi.org/10.1103/PhysRevLett.120.023201} {\bibfield  {journal}
  {\bibinfo  {journal} {Phys. Rev. Lett.}\ }\textbf {\bibinfo {volume} {120}},\
  \bibinfo {pages} {023201} (\bibinfo {year} {2018})}\BibitemShut {NoStop}%
\bibitem [{\citenamefont {Sackett}\ \emph {et~al.}(2000)\citenamefont
  {Sackett}, \citenamefont {Kielpinski}, \citenamefont {King}, \citenamefont
  {Langer}, \citenamefont {Meyer}, \citenamefont {Myatt}, \citenamefont {Rowe},
  \citenamefont {Turchette}, \citenamefont {Itano}, \citenamefont {Wineland},\
  and\ \citenamefont {Monroe}}]{Sackett2000}%
  \BibitemOpen
  \bibfield  {author} {\bibinfo {author} {\bibfnamefont {C.~A.}\ \bibnamefont
  {Sackett}}, \bibinfo {author} {\bibfnamefont {D.}~\bibnamefont {Kielpinski}},
  \bibinfo {author} {\bibfnamefont {B.~E.}\ \bibnamefont {King}}, \bibinfo
  {author} {\bibfnamefont {C.}~\bibnamefont {Langer}}, \bibinfo {author}
  {\bibfnamefont {V.}~\bibnamefont {Meyer}}, \bibinfo {author} {\bibfnamefont
  {C.~J.}\ \bibnamefont {Myatt}}, \bibinfo {author} {\bibfnamefont
  {M.}~\bibnamefont {Rowe}}, \bibinfo {author} {\bibfnamefont {Q.~A.}\
  \bibnamefont {Turchette}}, \bibinfo {author} {\bibfnamefont {W.~M.}\
  \bibnamefont {Itano}}, \bibinfo {author} {\bibfnamefont {D.~J.}\ \bibnamefont
  {Wineland}},\ and\ \bibinfo {author} {\bibfnamefont {C.}~\bibnamefont
  {Monroe}},\ }\href {https://doi.org/10.1038/35005011} {\bibfield  {journal}
  {\bibinfo  {journal} {Nature}\ }\textbf {\bibinfo {volume} {404}},\ \bibinfo
  {pages} {256} (\bibinfo {year} {2000})}\BibitemShut {NoStop}%
\bibitem [{\citenamefont {Kim}\ \emph {et~al.}(2009)\citenamefont {Kim},
  \citenamefont {Chang}, \citenamefont {Islam}, \citenamefont {Korenblit},
  \citenamefont {Duan},\ and\ \citenamefont {Monroe}}]{Kim2009}%
  \BibitemOpen
  \bibfield  {author} {\bibinfo {author} {\bibfnamefont {K.}~\bibnamefont
  {Kim}}, \bibinfo {author} {\bibfnamefont {M.-S.}\ \bibnamefont {Chang}},
  \bibinfo {author} {\bibfnamefont {R.}~\bibnamefont {Islam}}, \bibinfo
  {author} {\bibfnamefont {S.}~\bibnamefont {Korenblit}}, \bibinfo {author}
  {\bibfnamefont {L.-M.}\ \bibnamefont {Duan}},\ and\ \bibinfo {author}
  {\bibfnamefont {C.}~\bibnamefont {Monroe}},\ }\href
  {https://doi.org/10.1103/PhysRevLett.103.120502} {\bibfield  {journal}
  {\bibinfo  {journal} {Phys. Rev. Lett.}\ }\textbf {\bibinfo {volume} {103}},\
  \bibinfo {pages} {120502} (\bibinfo {year} {2009})}\BibitemShut {NoStop}%
\bibitem [{\citenamefont {Manning}(2014)}]{ManningThesis}%
  \BibitemOpen
  \bibfield  {author} {\bibinfo {author} {\bibfnamefont {T.~A.}\ \bibnamefont
  {Manning}},\ }\emph {\bibinfo {title} {Quantum Information Processing with
  Trapped Ion Chains}},\ \href@noop {} {Ph.D. thesis},\ \bibinfo  {school}
  {University of Maryland, College Park} (\bibinfo {year} {2014})\BibitemShut
  {NoStop}%
\bibitem [{\citenamefont {Figgatt}\ \emph {et~al.}(2019)\citenamefont
  {Figgatt}, \citenamefont {Ostrander}, \citenamefont {Linke}, \citenamefont
  {Landsman}, \citenamefont {Zhu}, \citenamefont {Maslov},\ and\ \citenamefont
  {Monroe}}]{Figgatt2019}%
  \BibitemOpen
  \bibfield  {author} {\bibinfo {author} {\bibfnamefont {C.}~\bibnamefont
  {Figgatt}}, \bibinfo {author} {\bibfnamefont {A.}~\bibnamefont {Ostrander}},
  \bibinfo {author} {\bibfnamefont {N.~M.}\ \bibnamefont {Linke}}, \bibinfo
  {author} {\bibfnamefont {K.~A.}\ \bibnamefont {Landsman}}, \bibinfo {author}
  {\bibfnamefont {D.}~\bibnamefont {Zhu}}, \bibinfo {author} {\bibfnamefont
  {D.}~\bibnamefont {Maslov}},\ and\ \bibinfo {author} {\bibfnamefont
  {C.}~\bibnamefont {Monroe}},\ }\href
  {https://doi.org/10.1038/s41586-019-1427-5} {\bibfield  {journal} {\bibinfo
  {journal} {Nature}\ }\textbf {\bibinfo {volume} {572}},\ \bibinfo {pages}
  {368} (\bibinfo {year} {2019})}\BibitemShut {NoStop}%
\bibitem [{Note1()}]{Note1}%
  \BibitemOpen
  \bibinfo {note} {The laser power required to perform the unbalanced Gaussian
  MS gate is significantly higher since there is only strong contribution from
  one motional mode instead of two. As such, we are limited to a detuning of
  -25.3\protect \,kHz to stay within the bounds of our laser system. We also
  use the center ion and an edge ion in the unbalanced case to take advantage
  of the stronger zig-zag coupling from the center ion.}\BibitemShut {Stop}%
\bibitem [{\citenamefont {Egan}(2021)}]{EganThesis}%
  \BibitemOpen
  \bibfield  {author} {\bibinfo {author} {\bibfnamefont {L.~N.}\ \bibnamefont
  {Egan}},\ }\emph {\bibinfo {title} {Scaling Quantum Computers with Long
  Chains of Trapped Ions}},\ \href@noop {} {Ph.D. thesis},\ \bibinfo  {school}
  {University of Maryland, College Park} (\bibinfo {year} {2021})\BibitemShut
  {NoStop}%
\bibitem [{\citenamefont {Brown}\ \emph {et~al.}(2021)\citenamefont {Brown},
  \citenamefont {Chiaverini}, \citenamefont {Sage},\ and\ \citenamefont
  {H{\"a}ffner}}]{BrownMaterials2021}%
  \BibitemOpen
  \bibfield  {author} {\bibinfo {author} {\bibfnamefont {K.~R.}\ \bibnamefont
  {Brown}}, \bibinfo {author} {\bibfnamefont {J.}~\bibnamefont {Chiaverini}},
  \bibinfo {author} {\bibfnamefont {J.~M.}\ \bibnamefont {Sage}},\ and\
  \bibinfo {author} {\bibfnamefont {H.}~\bibnamefont {H{\"a}ffner}},\ }\href
  {https://doi.org/10.1038/s41578-021-00292-1} {\bibfield  {journal} {\bibinfo
  {journal} {Nature Reviews Materials}\ }\textbf {\bibinfo {volume} {6}},\
  \bibinfo {pages} {892} (\bibinfo {year} {2021})}\BibitemShut {NoStop}%
\bibitem [{\citenamefont {Lin}\ \emph {et~al.}(2009)\citenamefont {Lin},
  \citenamefont {Zhu}, \citenamefont {Islam}, \citenamefont {Kim},
  \citenamefont {Chang}, \citenamefont {Korenblit}, \citenamefont {Monroe},\
  and\ \citenamefont {Duan}}]{Lin2009}%
  \BibitemOpen
  \bibfield  {author} {\bibinfo {author} {\bibfnamefont {G.-D.}\ \bibnamefont
  {Lin}}, \bibinfo {author} {\bibfnamefont {S.-L.}\ \bibnamefont {Zhu}},
  \bibinfo {author} {\bibfnamefont {R.}~\bibnamefont {Islam}}, \bibinfo
  {author} {\bibfnamefont {K.}~\bibnamefont {Kim}}, \bibinfo {author}
  {\bibfnamefont {M.-S.}\ \bibnamefont {Chang}}, \bibinfo {author}
  {\bibfnamefont {S.}~\bibnamefont {Korenblit}}, \bibinfo {author}
  {\bibfnamefont {C.}~\bibnamefont {Monroe}},\ and\ \bibinfo {author}
  {\bibfnamefont {L.-M.}\ \bibnamefont {Duan}},\ }\href
  {https://doi.org/10.1209/0295-5075/86/60004} {\bibfield  {journal} {\bibinfo
  {journal} {{EPL} (Europhysics Letters)}\ }\textbf {\bibinfo {volume} {86}},\
  \bibinfo {pages} {60004} (\bibinfo {year} {2009})}\BibitemShut {NoStop}%
\bibitem [{\citenamefont {Johanning}(2016)}]{Johanning2016}%
  \BibitemOpen
  \bibfield  {author} {\bibinfo {author} {\bibfnamefont {M.}~\bibnamefont
  {Johanning}},\ }\href {https://doi.org/10.1007/s00340-016-6340-0} {\bibfield
  {journal} {\bibinfo  {journal} {Applied Physics B}\ }\textbf {\bibinfo
  {volume} {122}},\ \bibinfo {pages} {71} (\bibinfo {year} {2016})}\BibitemShut
  {NoStop}%
\bibitem [{\citenamefont {Xie}\ \emph {et~al.}(2017)\citenamefont {Xie},
  \citenamefont {Zhang}, \citenamefont {Ou}, \citenamefont {Chen},
  \citenamefont {Zhang}, \citenamefont {Wu}, \citenamefont {Wu},\ and\
  \citenamefont {Chen}}]{Xie2017}%
  \BibitemOpen
  \bibfield  {author} {\bibinfo {author} {\bibfnamefont {Y.}~\bibnamefont
  {Xie}}, \bibinfo {author} {\bibfnamefont {X.}~\bibnamefont {Zhang}}, \bibinfo
  {author} {\bibfnamefont {B.}~\bibnamefont {Ou}}, \bibinfo {author}
  {\bibfnamefont {T.}~\bibnamefont {Chen}}, \bibinfo {author} {\bibfnamefont
  {J.}~\bibnamefont {Zhang}}, \bibinfo {author} {\bibfnamefont
  {C.}~\bibnamefont {Wu}}, \bibinfo {author} {\bibfnamefont {W.}~\bibnamefont
  {Wu}},\ and\ \bibinfo {author} {\bibfnamefont {P.}~\bibnamefont {Chen}},\
  }\href {https://doi.org/10.1103/PhysRevA.95.032341} {\bibfield  {journal}
  {\bibinfo  {journal} {Phys. Rev. A}\ }\textbf {\bibinfo {volume} {95}},\
  \bibinfo {pages} {032341} (\bibinfo {year} {2017})}\BibitemShut {NoStop}%
\bibitem [{\citenamefont {Crain}\ \emph {et~al.}(2014)\citenamefont {Crain},
  \citenamefont {Mount}, \citenamefont {Baek},\ and\ \citenamefont
  {Kim}}]{crain:2014}%
  \BibitemOpen
  \bibfield  {author} {\bibinfo {author} {\bibfnamefont {S.}~\bibnamefont
  {Crain}}, \bibinfo {author} {\bibfnamefont {E.}~\bibnamefont {Mount}},
  \bibinfo {author} {\bibfnamefont {S.}~\bibnamefont {Baek}},\ and\ \bibinfo
  {author} {\bibfnamefont {J.}~\bibnamefont {Kim}},\ }\href
  {https://doi.org/10.1063/1.4900754} {\bibfield  {journal} {\bibinfo
  {journal} {Applied Physics Letters}\ }\textbf {\bibinfo {volume} {105}},\
  \bibinfo {pages} {181115} (\bibinfo {year} {2014})}\BibitemShut {NoStop}%
\bibitem [{\citenamefont {Fang}\ \emph {et~al.}(2022)\citenamefont {Fang},
  \citenamefont {Wang}, \citenamefont {Huang}, \citenamefont {Brown},\ and\
  \citenamefont {Kim}}]{fang:2022}%
  \BibitemOpen
  \bibfield  {author} {\bibinfo {author} {\bibfnamefont {C.}~\bibnamefont
  {Fang}}, \bibinfo {author} {\bibfnamefont {Y.}~\bibnamefont {Wang}}, \bibinfo
  {author} {\bibfnamefont {S.}~\bibnamefont {Huang}}, \bibinfo {author}
  {\bibfnamefont {K.~R.}\ \bibnamefont {Brown}},\ and\ \bibinfo {author}
  {\bibfnamefont {J.}~\bibnamefont {Kim}},\ }\bibfield  {journal} {\bibinfo
  {journal} {arXiv}\ }\href {https://doi.org/10.48550/ARXIV.2206.02703}
  {10.48550/ARXIV.2206.02703} (\bibinfo {year} {2022})\BibitemShut {NoStop}%
\end{thebibliography}%

\end{document}